\documentclass[12pt, draftclsnofoot, onecolumn]{IEEEtran}
\usepackage[font={small}]{caption}
\usepackage{color}
\usepackage{graphicx}
\usepackage{epstopdf}
\usepackage{cite}
\usepackage{subfig}
\usepackage{amsthm}
\usepackage{amssymb}

\usepackage{epsfig}
\usepackage{amsfonts}
\usepackage{epstopdf}
\usepackage[]{algorithmic}
\usepackage[]{algorithm2e}
\usepackage{listings}
\usepackage{fancybox}
\usepackage{epsfig}
\usepackage{mathtools}
\usepackage{balance}
\usepackage[]{algorithm2e}

\usepackage{amsmath}
\ifCLASSINFOpdf
\else
\fi
\newcommand{\comm}[1]{}
\newcommand{\Nrf}{N_{\mathrm{RF}}}

\allowdisplaybreaks

\begin{document}
\title{Message passing-based link configuration in\\ short range millimeter wave systems}
\author{{\IEEEauthorblockN{Nitin Jonathan Myers, {\it Student Member, IEEE}, Jarkko Kaleva, {\it Member, IEEE},\\   Antti T\"olli, {\it Senior Member, IEEE}, and Robert W. Heath Jr., {\it Fellow, IEEE}. }}
\thanks{ N. J. Myers (nitinjmyers@utexas.edu) and R. W. Heath Jr. (rheath@utexas.edu) are with the Wireless Networking and Communications  Group, The University of Texas at Austin, Austin,
TX 78712 USA. A. T\"olli (antti.tolli@oulu.fi) and J. Kaleva (jarkko.kaleva@oulu.fi) are with Centre for Wireless Communications, University of Oulu, P.O. Box 4500, FIN-90014, Finland. This research was supported by the National Science Foundation under grant numbers NSF-CNS-1702800 and NSF-CNS-1731658, and by the Academy of Finland under grant numbers 311741 and 318927 (6Genesis Flagship). The material in this paper will appear in part in the IEEE Signal Processing Advances in Wireless Communications (SPAWC) 2019 conference \cite{shortrange_spawc}.}}
\maketitle
\vspace{-2mm}
\begin{abstract}
Millimeter wave (mmWave) communication in typical wearable and data center settings is short range. As the distance between the transmitter and the receiver in short range scenarios can be comparable to the length of the antenna arrays, the common far field approximation for the channel may not be applicable. As a result, dictionaries that result in a sparse channel representation in the far field setting may not be appropriate for short distances. In this paper, we develop a novel framework to exploit the structure in short range mmWave channels. The proposed method splits the channel into several subchannels for which the far field approximation can be applied. Then, the structure within and across different subchannels is leveraged using message passing. We show how information about the antenna array geometry can be used to design message passing factors that incorporate structure across successive subchannels. Simulation results indicate that our framework can be used to achieve better beam alignment with fewer channel measurements when compared to standard compressed sensing-based techniques that do not exploit structure across subchannels.
\end{abstract}
\begin{IEEEkeywords} 
Short range communication, mm-Wave, hybrid beamforming, dynamic compressed sensing.
\end{IEEEkeywords}
\IEEEpeerreviewmaketitle
\section{Introduction}
\par Millimeter wave (mmWave) systems can support high data rates and enable emerging applications like augmented reality and virtual reality \cite{mmWavemotiv}. Integrating mmWave radios in small scale devices like wearables can be easier as large antenna arrays occupy a small footprint at mmWave when compared to lower frequency systems \cite{formfactor}. Challenges like cost and power consumption in mmWave radios can be addressed by using analog beamforming-based radio designs or low-resolution receivers \cite{heathoverview}. To support high data rates in mmWave systems, the wireless link between the transmitter and the receiver must be configured properly. 
\par Establishing the link between the transmitting and receiving radios can be challenging at mmWave due to hardware constraints \cite{heathoverview}. As typical mmWave beamforming systems have a large number of antennas and fewer radio frequency (RF) chains, conventional beam alignment techniques may require a lot of training overhead when applied to such systems \cite{11ad}.  Algorithms that exploit structure in mmWave channels, like sparsity in an appropriate dictionary or low rank, can be useful in mmWave settings \cite{heathoverview}. Structure-aware algorithms, which are based on compressed sensing (CS) or matrix completion, can learn the mmWave channel or perform beam alignment with fewer channel measurements when compared to standard brute-force search approaches \cite{kiranchannel,lowrank1}. As a result, structure-aware algorithms can enable rapid mmWave link configuration. Most sparsity- or low rank-aware algorithms, however, make the far field assumption to model the multiple-input multiple-output (MIMO) channel \cite{kiranchannel,lowrank1}. In typical mmWave wearable settings where the distance between the transmitting and the receiving radios can be comparable to the length of the antenna arrays, the far field approximation may cease to hold \cite{tsewicomm}. Therefore, it is necessary to develop new channel estimation or beam alignment techniques that are tailored to short range mmWave systems. 
\par In this paper, we propose a new framework to leverage structure in short range mmWave communication channels. Our framework assumes a subarray-based hybrid beamforming system \cite{heathoverview}, and is based on two key observations. First, the far field approximation can be applied for a subchannel although it may fail for a full channel. As a result, it is reasonable to assume that the subchannels have a sparse representation at mmWave, in an appropriate dictionary. Second, the channel entries across appropriately chosen subchannels are correlated if the antenna arrays corresponding to the subchannels are co-located. Our framework splits the MIMO channel into several subchannels to exploit structure within and across the subchannels. 
\par The correlation across the sparse subchannel representations can be used to regularize the channel estimation or beam alignment problems. Prior work on dynamic CS has shown how signal recovery techniques that account for both correlation and sparsity of a group of vectors can outperform standard CS that solves for the vectors independently \cite{vaswani2010modified, DCSAMP}. Our framework allows applying such dynamic CS techniques for channel estimation and beam alignment in short range settings. Dynamic CS-based channel estimation techniques were developed for far field systems in \cite{kiranchannel,dynamiccs1,dynamiccs2,dynamiccs3} to exploit structure along time, frequency or spatial dimensions. For example, the far field channel was split into a group of subchannels for compressive subspace estimation \cite{KiranBS}. The splitting in \cite{KiranBS} was performed along the frequency dimension to deal with beam squint in wideband systems. To the best of our knowledge, however, short range link configuration techniques that are based on dynamic CS have not been developed. Furthermore, the proposed geometry-aware message passing technique, which incorporates information about the antenna geometry, is novel. We summarize the main contributions of our work as follows.
\begin{itemize}
\item We model a short range mmWave MIMO channel as a collection of several subchannels that satisfy the far field approximation. We use an appropriate sparsifying dictionary for the subchannels, and show that the support and amplitude of the sparse coefficients exhibit smooth dynamics across subchannels. 
\item We show how dynamic compressed sensing-approximate message passing (DCS-AMP) \cite{DCSAMP}, a dynamic CS algorithm, can be used to leverage the sparsity and correlation in subchannels for channel estimation. We also provide a low complexity implementation of DCS-AMP by establishing an equivalence between subchannel estimation and dynamic magnetic resonance imaging (MRI) CS. 
\item We develop an alternative message passing technique that uses information about the antenna geometry and the range of the transceiver distance, for compressive beam alignment. The proposed algorithm solves for the angles-of-arrival in subarrays, and can exploit subchannel structure beyond sparsity and smooth variation in subchannel coefficients.
\item We evaluate the performance of DCS-AMP and the proposed geometry-aided message passing algorithm, for a short range mmWave scenario in a multi-user setting. Our results indicate that the DCS-AMP-based approach is useful in the low SNR regime, while the proposed algorithm achieves better beam alignment at high SNR.
\end{itemize}
\par The link configuration techniques based on DCS-AMP and geometry-aided message passing use a different pilot structure, and exploit subchannel dynamics in different ways. On the one hand, pilot transmissions in the DCS-AMP-based approach are only performed in the uplink, to estimate the MIMO channel. On the other hand, the proposed geometry-aided message passing-based method uses both uplink and downlink pilot transmissions, as in the IEEE 802.11ad standard \cite{11ad}. Furthermore, the geometry-based method does not intend to estimate the MIMO channel, but solves for the local angles-of-arrivals (AoAs) that are defined for each subarray. As a result, the proposed technique requires low complexity and limited feedback when compared to the DCS-AMP-based method. Our geometry-aided message passing algorithm, however, can only recover the line-of-sight (LoS) component, unlike the DCS-AMP-based approach that can also recover the non-LoS components in the channel. The DCS-AMP-based technique accounts for the smooth variation in subchannels by assuming that the subchannel coefficients follow a Gauss-Markov dynamics. The proposed geometry-aided message passing algorithm explicitly captures the variation in subchannels by using message passing factors that are computed using the array geometry.
\par Notation : $\mathbf{A}$ is a matrix, $\mathbf{a}$ is a column vector and $a, A$ denote scalars. Using this notation $\mathbf{A}^T,\overline{\mathbf{A}}$ and $\mathbf{A}^{\ast} $ represent the transpose, conjugate and conjugate transpose of $\mathbf{A}$. The matrix $[\mathbf{A};\mathbf{B}]$ is obtained by vertically stacking $\mathbf{A}$ and $\mathbf{B}$. The spectral norm and the Frobenius norm of $\mathbf{A}$ are denoted by $\Vert \mathbf{A}\Vert _{2}$ and $\Vert \mathbf{A}\Vert _{F}$. The scalar $a\left[m \right]$ denotes the $m^{\mathrm{th}}$ element of $\mathbf{a}$, and $\mathbf{A}\left(k,\ell\right)$ is the entry of $\mathbf{A}$ in the $k^{\mathrm{th}}$ row and ${\ell}^{\mathrm{th}}$ column. The $k^{\mathrm{th}}$ row of $\mathbf{A}$ is denoted as $\mathbf{A}(k,:)$. The matrix $|\mathbf{A}|$ contains the element-wise magnitude of $\mathbf{A}$. The symbol $\odot$ is used for the Hadamard product. The matrices $\mathbf{U}_N$ and $\mathbf{I}_{N}$ in $\mathbb{C}^{N \times N}$ denote the unitary Discrete Fourier Transform (DFT) matrix and the identity matrix. We use $\mathbf{e}_k$ to represent the $(k+1)^{\mathrm{th}}$ canonical basis vector. The set $\mathcal{I}_N=\{1,2,3,.\,.\,.,N\}$. $\mathcal{N}_c(\mu , v_a)$ denotes the complex Gaussian distribution with mean $\mu$ and variance $v_a$. We define $\mathsf{j}=\sqrt{-1}$.
\section{System and channel model}
\par In this section, we explain the system and channel model in a short range mmWave setting. For simplicity of exposition, we assume a point-to-point link in a narrowband setting. The simulation results in Section \ref{sec:simulations} consider a more complex multi-user scenario.
\par
\begin{figure}[h!]
\vspace{-1mm}
\centering
\includegraphics[width=0.685 \linewidth]{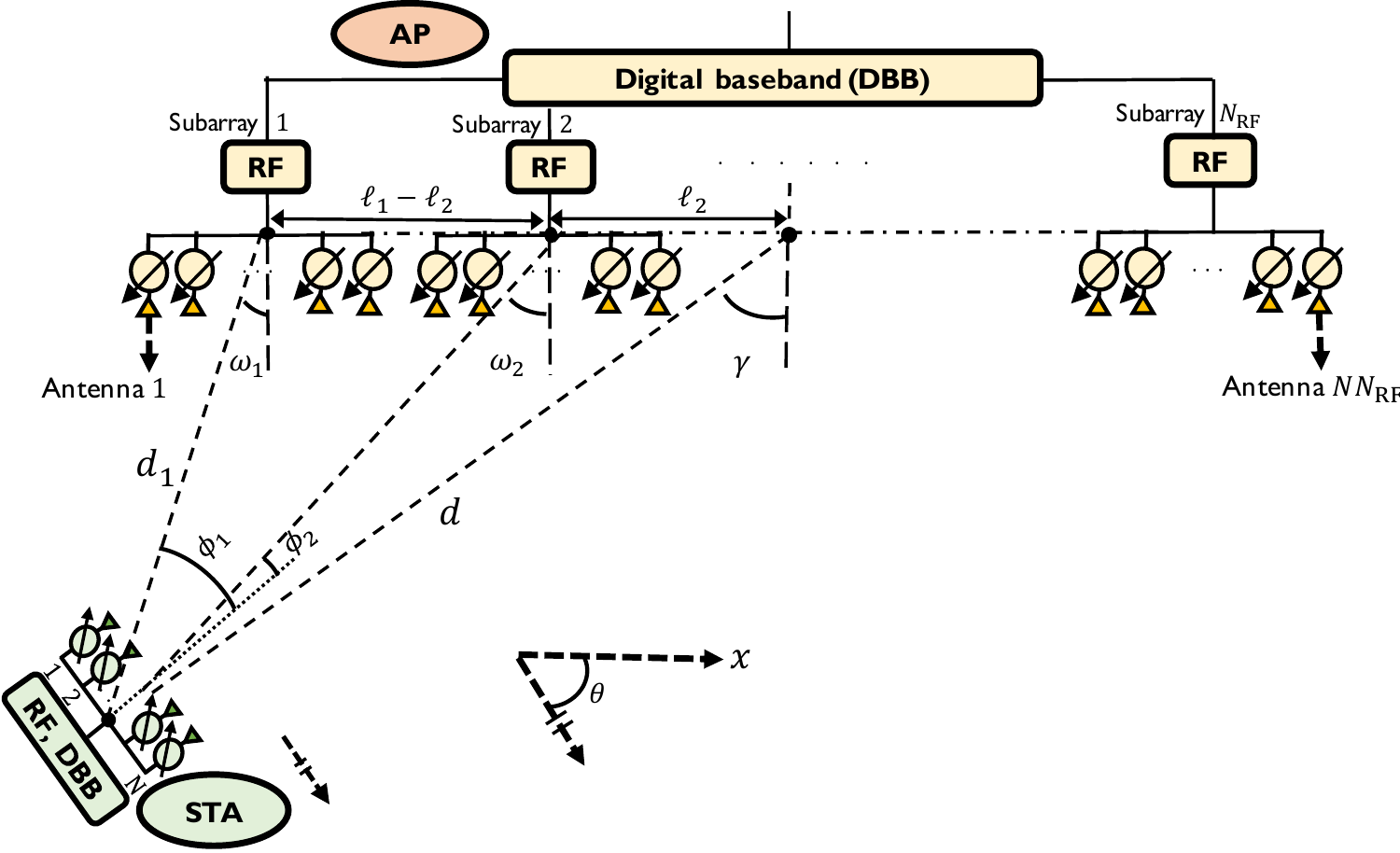}
  \caption{A short range line-of-sight mmWave communication system with linear arrays at both the AP and the STA. The AP and the STA are equipped with the hybrid and the analog beamforming architectures that use phase shifters.}
  \label{fig:architect}
\end{figure}

 We consider a mmWave setting in Fig.~\ref{fig:architect} where the access point (AP) has a subarray-based hybrid beamforming architecture with $\Nrf$ RF chains. Each of the $\Nrf$ subarrays at the AP is a half-wavelength spaced uniform linear array (ULA) with $N$ antennas. The $N$ antennas of a subarray are connected to a single RF chain through phase shifters. Therefore, each subarray consists of a single phased array. We use $\lambda$ to denote the wavelength corresponding to the mmWave carrier frequency. The resolution of all the $N\Nrf$ phase shifters at the AP is assumed to be $q$ bits. The use of low resolution phase shifters results in a lower power consumption, and introduces new challenges for beam alignment. These challenges can be addressed by a careful design of codebooks \cite{FALP}. The alphabet for phase shifts is defined as $\mathbb{Q}=\{e^{\mathsf{j}2 \pi \ell/2^q}/ \sqrt{N}  : \ell \in \mathcal{I}_{2^q} \}$. The spacing between consecutive subarrays is assumed to be constant; this spacing can be larger than $\lambda/2$. The antennas at the AP are placed along the horizontal axis and occupy a length of $L_{\mathrm{AP}}$ units. Consider a single station (STA) in Fig.~\ref{fig:architect} equipped with a half-wavelength spaced ULA of $N$ elements. We assume that the antenna arrays at the AP and the STA operate at the same polarization and are coplanar. Let $d$ denote the distance between the midpoints of the antenna arrays at the AP and the STA, and $L_{\mathrm{STA}}$ denote the length of the antenna array at the STA. We define the angle $\gamma$ as the orientation of the STA relative to the normal of the AP array, and $\theta$ as the angle made by the STA array with the horizontal. The ULA at the STA is equipped with phase shifters and a single RF chain. Our framework can also be extended to other antenna architectures like uniform planar arrays, by using appropriate array response vectors in the formulation. 
\par Now, we describe the system model used to obtain uplink channel measurements. We assume a narrowband channel for simplicity. Extending our framework to incorporate frequency selectiveness of channels is an interesting research direction. We consider pilot transmissions in the uplink as the AP may have better computational capabilities than the STA, and can acquire $\Nrf$ channel measurements in parallel. Let $\mathbf{H}_k \in \mathbb{C}^{N \times N}$ be the baseband equivalent of the short range MIMO channel matrix between the STA and the $k^{\mathrm{th}}$ subarray at the AP. The matrix $\mathbf{H}_k$ is defined as the $k^{\mathrm{th}}$ subchannel in the MIMO system. We define the full MIMO channel matrix as $\mathbf{H}\in \mathbb{C}^{N\Nrf \times N}$. The channel $\mathbf{H}$ is defined as $\mathbf{H}=[\mathbf{H}_1 ; \mathbf{H}_2 ; \cdots ; \mathbf{H}_{\Nrf}]$. Channel measurements in the MIMO system are acquired by applying beam training vectors to the phased arrays at the STA and the AP. Let ${\mathbf{f}}[m] \in \mathbb{Q}^{N}$ be the transmit beam training vector applied to the phased array at the STA in the $m^{\mathrm{th}}$ training slot. During this slot, the AP acquires channel measurements by applying ${\mathbf{w}_k}[m] \in \mathbb{Q}^{N}$ as the receive beam training vector to its $k^{\mathrm{th}}$ subarray. The channel measurement acquired by the $k^{\mathrm{th}}$ subarray in the $m^{\mathrm{th}}$ slot is defined as
\begin{equation}
\label{eq:sysmodel}
{y}_k[m]=\mathbf{w}^{T}_k[m]\mathbf{H}_k{\mathbf{f}[m]}+v_k[m] , \;\;\; \forall k \in \{1,2,...\Nrf\},
\end{equation} 
 where $v_k[m] \sim \mathcal{N}_c(0,\sigma^2)$ is additive white Gaussian noise. It can be noticed from \eqref{eq:sysmodel} that $\Nrf$ projections of the channel matrix can be obtained in a given training slot. As $\mathbf{H}$ consists of $\Nrf N^2$ entries, beam alignment through conventional MIMO channel estimation requires $N^2$ training slots. In this paper, we show that a reasonable approximation of the MIMO channel $\mathbf{H}$ can be estimated in far fewer than $N^2$ slots by exploiting structure in the short range mmWave communication channel. The channel approximation obtained is then used to configure the antenna arrays at the AP and the STA.
\par To explain the key ideas underlying our framework, we consider a geometric LoS channel model in a short range setting. The simulation results in Section~\ref{sec:simulations}, however, consider a more realistic short range scenario that also includes non-LoS components. We use  $d_{i j}$ to denote the distance between the $i ^{\mathrm{th}}$ antenna element at the AP and the $j^{\mathrm{th}}$ element at the STA. Here, $i \in \mathcal{I}_{N \Nrf}$ and $j \in \mathcal{I}_N$. The discrete time baseband equivalent of the MIMO channel matrix can be expressed as \cite{tsewicomm}
\begin{equation}
\mathbf{H}(i,j)=\frac{\lambda}{4 \pi d_{ij}}e^{-\mathsf{j}2\pi d_{ij}/ \lambda}.
\label{eq:shortchan}
\end{equation} 
The channel matrix with entries defined by \eqref{eq:shortchan} can be approximated as a rank-one matrix when $d$ is significantly larger than the length of the antenna arrays at the AP and the STA, i.e., $d \gg \mathrm{max}(L,(N-1)\lambda/2)$ \cite{rappwireless}. In settings where $d$ can be comparable to the length of the antenna array, the full channel matrix, i.e., $\mathbf{H}$, cannot always be approximated as a rank-one matrix.
\par Now, we show how subchannels exhibit a rank-one characteristic when compared to the full channel for reasonably short distances. To explain our argument, we define an energy metric that measures the rank-one nature of a channel. We define $E_\mathrm{F}(d)=\Vert \mathbf{H} \Vert^2_2/\Vert \mathbf{H} \Vert^2_F$ and $E_\mathrm{S}(d)=\sum^{\Nrf}_{k=1}\Vert \mathbf{H}_k \Vert^2_2/(\Vert \mathbf{H}_k \Vert^2_F \Nrf)$ as the energy metrics for the full channel and the subchannels. While $E_\mathrm{F}(d)$ represents the fraction of energy contained in the rank-one approximation of the full channel, $E_\mathrm{S}(d)$ denotes the average of the fractions computed for each subchannel. 
\begin{figure}[h!]
\vspace{-1mm}
\centering
\includegraphics[trim=1.5cm 6.25cm 2cm 7.5cm,clip=true,width=0.55 \textwidth]{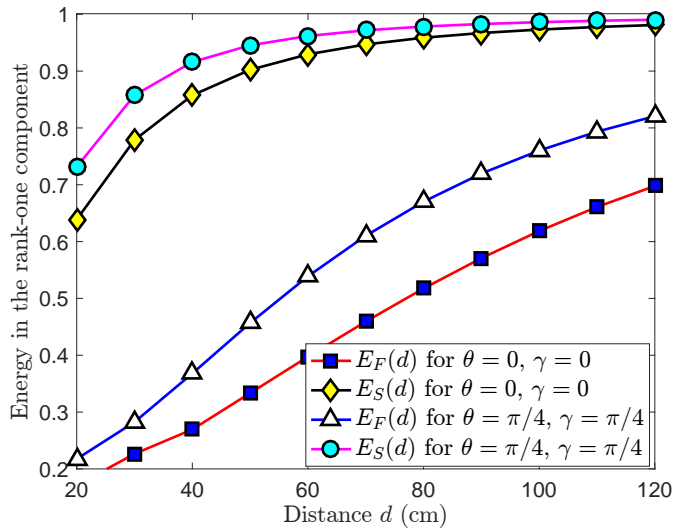}
  \caption{The plot shows the variation in the energy metric with the AP-STA distance, i.e., $d$. The parameters of the antenna arrays were set as $L_{\mathrm{AP}}=20\,\mathrm{cm}$, $L_{\mathrm{STA}}=4\,\mathrm{cm}$, $N=16$, and $\Nrf=4$. Here, $\lambda= 5 \, \mathrm{mm}$. For a reasonably small $d$, subchannels can be well approximated as rank-one matrices when compared to full channel matrix.}
  \label{fig:structdist}
\end{figure}
For the short range communication setting in Fig.~\ref{fig:architect}, it can be observed from Fig.~\ref{fig:structdist} that the energy metric increases as a function of distance $d$. The observation is consistent with the fact that LoS MIMO channels can be well approximated as rank-one in the far field setting. An important observation from Fig.~\ref{fig:structdist} is that the energy metric for subchannels, i.e., $E_{\mathrm{S}}(d)$, is higher than that for the full channel. Therefore, subchannels can be better approximated as rank-one matrices when compared to the full channel. The approximation is reasonable as the far field assumption is likely to be valid for arrays of shorter length, i.e., subarrays, when compared to the full array.
\par The array response vectors in a subchannel can be modeled as Vandermonde vectors when the rank-one approximation is valid for the subchannel \cite{rappwireless}. For the half wavelength-spaced ULA, we define the array response vector of dimension $N$ as 
\begin{equation}
\mathbf{a}_{_{N}}(\Delta)=[1,e^{-\mathsf{j} \pi \mathrm{sin}\Delta},e^{-\mathsf{j}2\pi \mathrm{sin}\Delta},\,.\,.\, ,e^{-\mathsf{j}(N-1)\pi \mathrm{sin}\Delta}]^T.
\end{equation}
We use $\alpha_k$ to denote the path gain of the ray between the STA and the $k^{\mathrm{th}}$ subarray at the AP. The angle of departure and the angle of arrival of the ray, relative to the boresight, are denoted by $\phi_{k}$ and $\omega_k$. Note that a unique angle of arrival and departure can be defined for an array only when the far field approximation, equivalently the rank-one approximation, is valid. The angles $\{\omega_k\}_{k=1}^{\Nrf}$ are called the local AoAs, as they are defined for subarrays or sections of the AP array. Under the rank-one approximation of each subchannel, we can write $\mathbf{H}_k \approx \alpha_k \mathbf{a}_{_{N}}(\omega_k)\mathbf{a}^T_{_{N}}(\phi_k)$ for every $k \in \mathcal{I}_{\Nrf}$. Due to the Vandermonde structure of the array response vectors, $\mathbf{H}_k$ is compressible in the 2D-DFT dictionary. We use $\mathbf{X}_k$ to denote the inverse 2D-DFT of $\mathbf{H}_k$ such that 
\begin{equation}
\mathbf{H}_k= \mathbf{U}_N \mathbf{X}_k \mathbf{U}_N.
\end{equation} 
The matrix $\mathbf{X}_k$, defined as the beamspace subchannel, is approximately sparse \cite{heathoverview}. It is not exactly sparse as the angles of departure and arrival may not be aligned with those corresponding to the DFT dictionary \cite{kiranchannel}. 
\par
\begin{figure}[h!]
\vspace{-4mm}
\centering
\includegraphics[trim=2.5cm 6.5cm 2cm 6.8cm,clip=true,width=0.48 \textwidth]{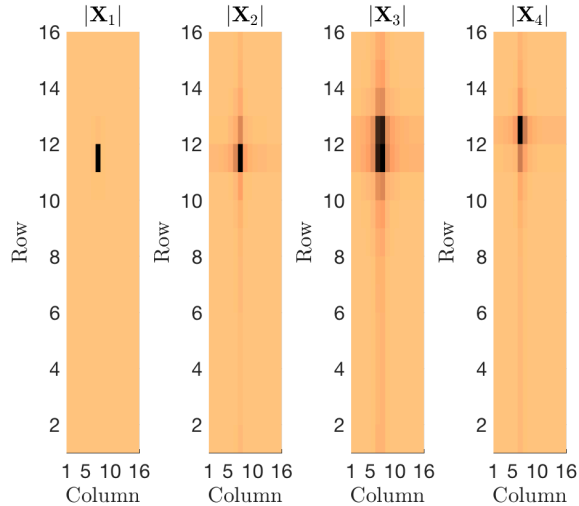}
  \caption{The plot shows the smooth variation in the locations of the non-zero beamspace coefficients across different subarrays. Here, $\lambda= 5 \, \mathrm{mm}$. The parameters of the antenna arrays were set as $L_{\mathrm{AP}}=20\,\mathrm{cm}$, $N=16$, $\Nrf=4$, $d=80\,\mathrm{cm}$, $\gamma=0$ and $\theta=\pi/4$. }
  \vspace{-2mm}
  \label{fig:spatcorr}
\end{figure}

 Now, we argue that the non-zero coefficients in $\{\mathbf{X}_k\}_{k=1}^{\Nrf}$ can exhibit smooth dynamics across the subchannel index $k$ in short range settings. The angles of arrival and departure determine the support of the non-zero coefficients in the beamspace subchannel \cite{beamsparse}. As the subarrays at the AP are co-located, the angles of departure and arrival, i.e., $\{ \phi_k \}_{k=1}^{\Nrf}$ and $\{\omega_k \}_{k=1}^{\Nrf}$, vary slowly with $k$. Therefore, the support of the non-zero coefficients in $\mathbf{X}_k$ drifts slowly with $k$. For the extreme case when $d \gg \mathrm{max}(L,(N-1)\lambda/2)$, it can be shown that the drift is negligible. An illustration of the dynamics of the coefficients in $\{\mathbf{X}_k\}_{k=1}^{\Nrf}$  is provided in Fig.~\ref{fig:spatcorr} for the short range communication setting in Fig.~\ref{fig:architect}. The beamspace subchannels in the short range system can be estimated from sub-Nyquist channel measurements by exploiting the sparse and correlated nature of subchannels.
\section{Short range channel estimation using dynamic compressed sensing}
In this section, we transform the beamspace subchannel estimation problem into an equivalent masked beamspace subchannel estimation problem by using Zadoff-Chu sequences in the antenna domain \cite{swiftlink}. Under such a transformation, we show that short range subchannel estimation is exactly equivalent to magnetic resonance imaging of a sequence of correlated angiogram images \cite{DCSAMP}. Then, we show how DCS-AMP algorithm \cite{DCSAMP} can be used to estimate the masked beamspace subchannels with $M \ll N^2$ pilot transmissions. 
\subsection{Zadoff-Chu-based training and the spectral mask concept} \label{sec:ZCacquisition}
\par Zadoff-Chu (ZC) sequences have been extensively applied along the time and frequency dimensions in wireless systems due to their constant amplitude and zero autocorrelation (CAZAC) properties \cite{zadoffchu}. Recent work has demonstrated the advantages of using ZC sequences over the common random phase shift sequences for CS-based beam alignment \cite{swiftlink,ZCglobecom,ZCCS_recent}. In this section, we use ZC sequences along the spatial dimension for compressive channel estimation in the subarray-based hybrid beamforming architecture shown in Fig.~\ref{fig:architect}. We define $\mathbf{z} \in \mathbb{C}^N$ as a unit norm ZC sequence of root $t \in \mathcal{I}_N$. The root $t$ must be co-prime with $N$. The $n^{\mathrm{th}}$ element of $\mathbf{z}$ is defined as \cite{zadoffchu}
\begin{align}
\label{eq:ZCdefn}
z\left[n\right]&=\begin{cases}
\begin{array}{c}
\frac{1}{\sqrt{N}}\mathrm{exp}\left(\mathsf{j}\frac{\pi t n\left(n+1\right)}{N}\right),\,\,\,\,\,\mathrm{if}\,N\,\mathrm{is\,odd}\\
\frac{1}{\sqrt{N}}\mathrm{exp}\left(\mathsf{j}\frac{\pi t n^{2}}{N}\right),\,\,\,\,\,\,\,\,\,\,\,\,\,\,\,\,\mathrm{if}\,N\,\mathrm{is\,even}
\end{array}
\end{cases}.
\end{align}
The constant amplitude of the entries of the ZC sequence allows realizing $\mathbf{z}$ in phased arrays. The constant amplitude of the DFT of $\mathbf{z}$ allows efficient compressed sensing-based beam alignment of 2D-DFT sparse channels \cite{ZCglobecom}. For an extensive treatment on the optimality of ZC-based channel acquisition for CS-based beam alignment, we refer the interested reader to \cite{ZCglobecom} and \cite{ZCCS_recent}.
\par The channel measurements for DCS-AMP-based channel estimation are acquired by applying random circulant shifts of the ZC sequence to the phased arrays at the AP and the STA. For the $m^{\mathrm{th}}$ transmission in a sequence of $M$ uplink pilot transmissions, the STA applies a $c[m] \in \mathcal{I}_N$ circulantly shifted version of $\mathbf{z}$ to its phased array. At the same time, the AP uses an $r_k[m] \in \mathcal{I}_N$ circulantly shifted version of $\mathbf{z}$ as the beam training vector for its $k^{\mathrm{th}}$ subarray. The pair of ZC sequences used at the STA and the $k^{\mathrm{th}}$ subarray at the AP determine the channel measurement corresponding to $\mathbf{H}_k$. As the STA and an AP subarray can independently apply $N$ different circulant shifts to their phased array, a maximum of $N^2$ spatially distinct subchannel measurements can be obtained with the ZC-based training. In this paper, we focus on the sub-Nyquist setting, i.e., $M<N^2$ measurements are acquired per subchannel. For each $m$, the STA chooses $c[m] \in \mathcal{I}_N$ at random with replacement. The AP chooses $\{r_k[m]\}^{\Nrf}_{k=1}$ at random such that none of the coordinates in $\{(r_k[m],c[m])\}^{\Nrf}_{k=1}$ were sampled in the previous random draws. Each of the $\Nrf$ subchannels is projected onto a set of $M$ linearly independent matrices that are determined by the random sampling scheme.
\par Now, we define masked beamspace subchannels using the spectral mask concept in \cite{swiftlink}. We define $\mathbf{J}\in \mathbb{R}^{N \times N}$ as a circulant matrix with its first row as $\left(0,1,0,...,0\right)$. The subsequent rows of $\mathbf{J}$ are generated by circulantly shifting the previous row to the right by one unit. We define the $\ell$-circulant delay matrix as $\mathbf{J}_{\ell}=\mathbf{J}\mathbf{J}\cdot \cdot \cdot \mathbf{J} \,(\mathrm{ \ell \, times})$. The matrix $\mathbf{J}_0$ is defined as the $N \times N$ identity matrix. Using these definitions, it can be observed that the beam training vector used by the STA for the $m^{\mathrm{th}}$ pilot transmission is $\mathbf{f}[m]=\mathbf{J}_{c[m]}\mathbf{z}$.  Similarly, the AP applies $\mathbf{w}_k[m]=\mathbf{J}_{r_k[m]}\mathbf{z}$ to its $k^{\mathrm{th}}$ subarray, to acquire a channel measurement for the $m^{\mathrm{th}}$ pilot transmission. From the system model in \eqref{eq:sysmodel}, the channel measurement $y_k[m]$ can be expressed as
\begin{equation}
y_k[m]=(\mathbf{J}_{r_k[m]}\mathbf{z})^T \mathbf{U}_N \mathbf{X}_k \mathbf{U}_N \mathbf{J}_{c[m]}\mathbf{z} +v_k[m].
\label{eq:sysmodel_circ}
\end{equation}
We define $\boldsymbol{\Lambda}_{\mathbf{z}}=\mathrm{diag}(\sqrt{N}\mathbf{U}_N \mathbf{z})$ as a diagonal matrix that contains the scaled DFT of $\mathbf{z}$ along its diagonal. The matrix $\boldsymbol{\Lambda}_{\mathbf{z}}$ is unimodular along its diagonal by the zero autocorrelation property of the ZC sequence \cite{ZCglobecom}. For each $k$, the masked beamspace subchannel associated with $\mathbf{X}_k$ is defined as $\mathbf{S}_k=\boldsymbol{\Lambda}_{\mathbf{z}}\mathbf{X}_k\boldsymbol{\Lambda}_{\mathbf{z}}$. Using the spectral mask concept \cite{swiftlink}, \eqref{eq:sysmodel_circ} can be rewritten as 
\begin{equation}
y_k[m]=\mathbf{e}_{r_k[m]}^T \mathbf{U}_N \mathbf{S}_k \mathbf{U}_N \mathbf{e}_{c[m]}+v_k[m].
\label{eq:sysmodel_masked}
\end{equation}
The spectral mask concept \cite{swiftlink} shows that estimating the masked beamspace subchannel $\mathbf{S}_k$ is equivalent to estimating the true beamspace subchannel $\mathbf{X}_k$. To recover the sparse vector $\mathbf{s}_k=\mathrm{vec}(\mathbf{S}_k)$, the CS matrix corresponding to the linear model in \eqref{eq:sysmodel_masked} is defined as $\mathbf{A}_k \in \mathbb{C}^{M \times N^2}$, where 
\begin{equation}
\mathbf{A}_k (m,:)=  (\mathbf{e}^T_{c[m]}  \mathbf{U}_N) \otimes (\mathbf{e}_{r_k[m]}^T \mathbf{U}_N).
\label{eq:csmatrixdefn}
\end{equation}
The channel measurement $y_k[m]$ can be expressed using \eqref{eq:sysmodel_masked} and \eqref{eq:csmatrixdefn} as
\begin{equation}
y_k[m]=\mathbf{A}_k \mathbf{s}_k + v_k[m].
\label{eq:standardcsmodel}
\end{equation} 
From \eqref{eq:sysmodel_masked}, the channel measurement in \eqref{eq:sysmodel_circ} can be interpreted as a noisy version of the $(r_k[m],c[m])$ sample of the 2D-DFT of the masked subchannel $\mathbf{S}_k$. As a result, the CS problem for masked beamspace subchannel recovery is a partial 2D-DFT CS problem \cite{swiftlink}.  
\par The map between $\mathbf{X}_k=\boldsymbol{\Lambda}^{-1}_{\mathbf{z}}\mathbf{S}_k\boldsymbol{\Lambda}^{-1}_{\mathbf{z}}$ and $\mathbf{S}_k$ is perfectly conditioned and invertible, due to the unimodular nature of the diagonal of $\boldsymbol{\Lambda}_{\mathbf{z}}$. Specifically, $|\mathbf{S}_k(r,c)|=|\mathbf{X}_k(r,c)|$ for every $k$, $r$ and $c$. Therefore, the masked beamspace subchannel $\mathbf{S}_k$ is sparse with the same locations of sparsity as that of the beamspace subchannel $\mathbf{X}_k$. Furthermore, the dynamics across the masked beamspace subchannels is smooth as the coefficients in the beamspace subchannels exhibit smooth variation across the subchannel index. Due to the structure preserving nature of the spectral mask, it is reasonable to solve for the masked beamspace subchannels instead of the original beamspace subchannels. From a computational complexity perspective, solving for $\{\mathbf{S}_k\}_{k=1}^{\Nrf}$ from the channel measurements is advantageous because CS matrix multiplications corresponding to \eqref{eq:sysmodel_masked} can be efficiently implemented using the fast Fourier transform (FFT) \cite{swiftlink}. 
\par The problem of recovering masked beamspace subchannels is similar to that of recovering a sequence of temporally correlated angiogram images in magnetic resonance imaging (MRI) \cite{DCSAMP}. The sequence of sparse matrices $\{\mathbf{S}_k\}_{k=1}^{\Nrf}$ is analogous to the sequence of sparse angiogram images that are sampled across time. In both settings, the matrices exhibit smooth dynamics across the sequence index. Furthermore, image acquisition in MR-based angiogram imaging and the ZC-based channel acquisition obtain samples of the 2D-DFT of a sparse matrix. The equivalence between the two problems, and the success of DCS-AMP in recovering a sequence of smoothly varying MR images motivates its application for short range mmWave channel estimation.  
\subsection{Masked beamspace recovery using DCS-AMP}
In this section, we explain how the sequence $\{\mathbf{s}_k\}_{k=1}^{\Nrf}$, i.e., the vector versions of the masked beamspace subchannels, can be recovered from the channel measurements $\{\mathbf{y}_k\}_{k=1}^{\Nrf}$ using DCS-AMP \cite{DCSAMP}. Standard CS algorithms like approximate message passing (AMP) \cite{AMP}, when applied independently over the $\Nrf$ sub-problems, i.e., each sub-problem solves for a subchannel, do not leverage the smooth variation across $\{\mathbf{s}_k\}_{k=1}^{\Nrf}$ for channel estimation. The correlation across these vectors can be used to regularize the problem, and dynamic CS techniques like DCS-AMP \cite{DCSAMP} that account for such correlation can outperform standard AMP-based CS. 
\par
\begin{figure}[h!]
\centering
\includegraphics[trim=2.5cm 1cm 2.5cm 0.5cm,clip=true,width=0.55 \linewidth]{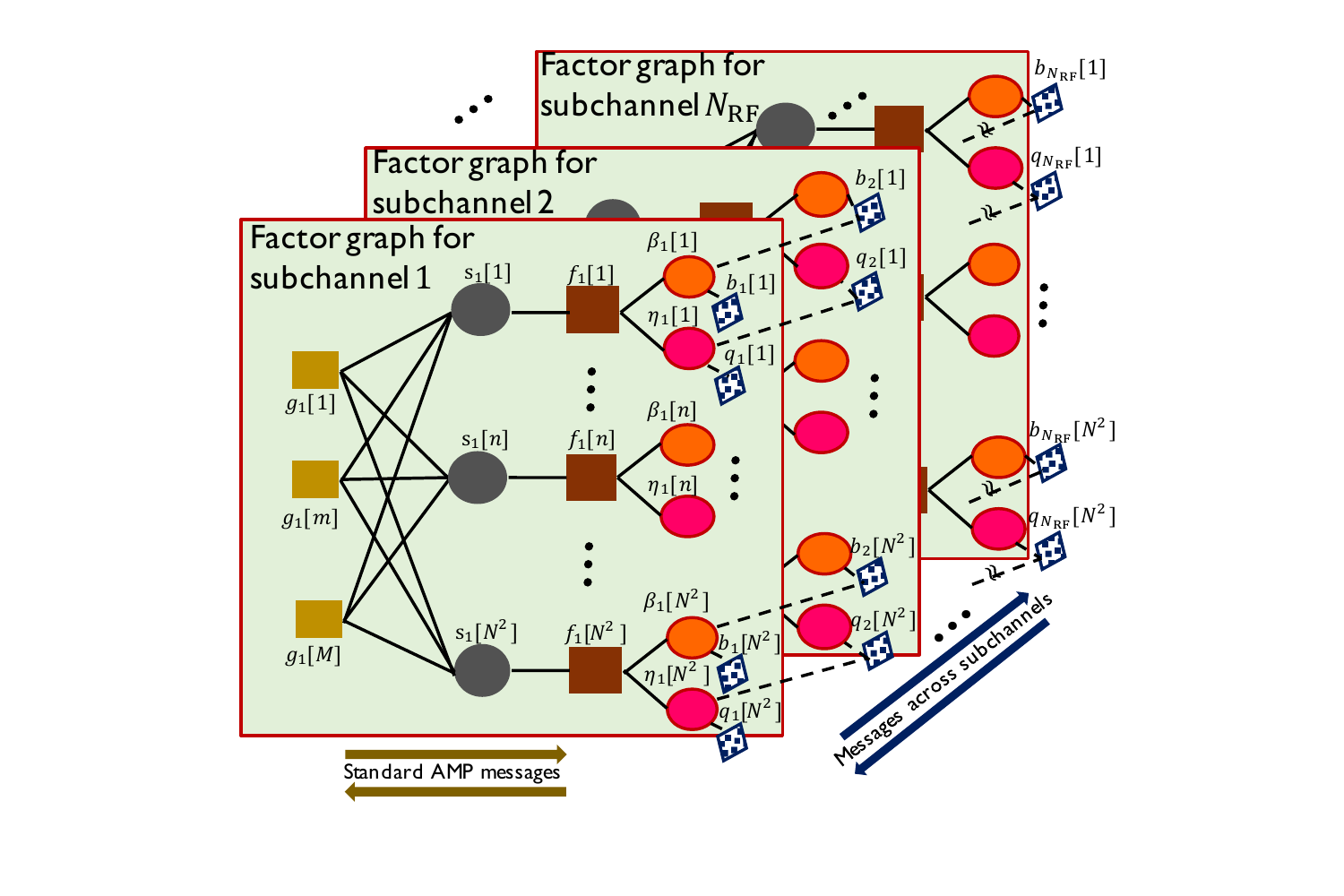}
  \caption{Factor graph for short range channel estimation using DCS-AMP.  In addition to standard AMP-based messages  for each subchannel, messages are also sent across different subchannels in DCS-AMP.}
  \label{fig:fact_graph}
\end{figure}

  In DCS-AMP \cite{DCSAMP}, the vectors $\{\mathbf{y}_k\}_{k=1}^{\Nrf}$ and  $\{\mathbf{s}_k\}_{k=1}^{\Nrf}$ are considered as the realizations of random vectors $\{\boldsymbol{\mathsf{y}}_k\}_{k=1}^{\Nrf}$ and  $\{\boldsymbol{\mathsf{s}}_k\}_{k=1}^{\Nrf}$. The random variables $\beta_k[n] \in \{0,1\}$ and $\eta_k[n] \in \mathbb{C}$ are used to model the support and amplitude of $\mathsf{s}_k[n]$, i.e., $\mathsf{s}_k[n]=\beta_k[n]\eta_k[n], \, \forall \, k, n$. The factor graph that illustrates the dependencies between the random variables is shown in Fig.~\ref{fig:fact_graph}. The elements of the random vectors are modeled using circular nodes, also called as variable nodes, in the factor graph. The rectangular nodes in Fig.~\ref{fig:fact_graph} are called the factors. Factors in Fig.~\ref{fig:fact_graph} are used to incorporate information about the random variables like sparsity and structural dependencies. For example, the dependency between $\mathsf{s}_k[n]$, $\beta_k[n]$ and $\eta_k[n]$ is captured using the factor $f_k[n]=\delta (\mathsf{s}_k[n]-\beta_k[n]\eta_k[n])$, where $\delta(\cdot)$ denotes the Dirac delta function \cite{DCSAMP}. The Dirac delta factor $f_k[n]$ ensures that the random variables satisfy $\mathsf{s}_k[n]=\beta_k[n]\eta_k[n]$. Each plane in the factor graph contains the variable nodes and factors corresponding to a particular subchannel. DCS-AMP performs information flows, i.e., probability distribution flows, within and across different planes to obtain subchannel estimates.
\par Channel measurements acquired by the subarrays provide information about the random vectors $\{\boldsymbol{\mathsf{s}}_k\}_{k=1}^{\Nrf}$. It can be observed from \eqref{eq:standardcsmodel} that $y_k[m]$ is a realization of $\mathbf{A}_k(m,:)\boldsymbol{\mathsf{s}}_k$, i.e., the projection of the random vector $\boldsymbol{\mathsf{s}}_k$. The factor corresponding to $y_k[m]$ is denoted by $g_k[m]$ in Fig.~\ref{fig:fact_graph}. The factors $\{g_k[m]\}_{k,m}$ ensure fidelity of the message passing solution to the observed channel measurements. It can be observed from \eqref{eq:sysmodel_masked} and \eqref{eq:standardcsmodel} that $y_k[m]$ depends on all the entries of $\mathbf{s}_k$ through a 2D-DFT equation. Therefore, $g_{k}[m]$, the factor corresponding to $y_k[m]$, is linked to all the $N^2$ elements of the random vector $\boldsymbol{\mathsf{s}}_k$ through a 2D-DFT. Specifically, $g_{k}[m]$ contains the likelihood function 
\begin{equation}
p(y_{k}[m]\,|\,\boldsymbol{\mathsf{s}}_{k})=\frac{1}{\sqrt{2\pi\sigma^{2}}}\, \mathrm{exp}\left(-\frac{|y_{k}[m]-\mathbf{A}_k(m,:)\boldsymbol{\mathsf{s}}_{k}|^{2}}{2\sigma^{2}}\right).
\end{equation}
The factor graph built until this point is applicable to any generic linear regression problem.
\par Now, we describe the key components in DCS-AMP that incorporate sparsity and structural dependencies in the beamspace subchannels. For every $n \in \mathcal{I}_{N^2}$, the support correlation across the subchannels is incorporated by assuming that $\{\beta_k[n]\}^{\Nrf}_{k=1}$ follow a binary Markov process \cite{DCSAMP}. The $N^2$ different random processes are assumed to be independent with the same transition probabilities that characterize a binary Markov process. The correlation in the amplitudes of the beamspace subchannels is taken into account by assuming that $\{\theta_k[n]\}^{\Nrf}_{k=1}$ follows a Gauss-Markov process \cite{DCSAMP}. Similar to the binary Markov case, the $N^2$ different Gauss-Markov processes are assumed to be independent with the same correlation and variance parameters. The factors corresponding to the $n^{\mathrm{th}}$ binary Markov process and the Gauss-Markov processed are modeled using $\{b_k[n]\}_{k=1}^{\Nrf}$ and $\{q_k[n]\}_{k=1}^{\Nrf}$ in the factor graph. DCS-AMP assumes that both the Markov processes are in steady state. The steady state probability that $\beta_k[n]=1$ is denoted by $\epsilon$. The scalars $\kappa$, $\zeta$ and $\rho$ are used to denote the correlation coefficient, the mean and the variance of the steady state Gauss-Markov process corresponding to an $\eta_k[n]$. As $\mathsf{s}_k[n]=\beta_k[n] \eta_k [n]$, it can be observed that $\mathsf{s}_k[n]$ is $0$ with a probability of  $1-\epsilon$, or is distributed as a Gaussian with mean $\zeta$ and variance $\rho$ with a probability of $\epsilon$. As a result, $\mathsf{s}_k[n]$ follows a Bernoulli-Gaussian (BG) prior, which is a reasonable distribution to model sparse signals. For detailed mathematical expressions of the factors in Fig.~\ref{fig:fact_graph}, we refer the interested reader to \cite{DCSAMP}. 
\par Similar to standard message passing, DCS-AMP performs iterative message flows between the variable nodes and the factors to estimate the marginal posteriors of $\{\mathsf{s}_{k}[n]\}_{k,n}$ for the channel measurements $\{y_{k}[m]\}_{k,m}$. We explain the sequence of message flows using $f_1[1]$. The factors $b_1[1]$ and $q_1[1]$ send their messages, i.e., probability distributions, to $f_1[1]$ through $\beta_1[1]$ and $\eta_1[1]$. Then, $f_1[1]$ combines the received messages to generate a BG distribution. Standard AMP iterations between $\{ g_1[m] \}^M_{m=1}$ and $\{ \mathsf{s}_1[n] \}^{N^2}_{n=1}$ are performed for subchannel $1$ using the BG distributions received from the factors $\{ f_1[n] \}^{N^2}_{n=1}$. At the end of these iterations, $f_1[1]$ receives Gaussian distributions from the factors $\{ g_1[m] \}^M_{m=1}$. These messages are appropriately combined and sent to the next layer, i.e., factor graph for subchannel 2, through $b_2[1]$ and $q_2[1]$. Then, message passing is performed in layer $2$. The process of message flows within and across the layers is stopped when the $\Nrf ^{\mathrm{th}}$ layer is updated. Such sequence of flows across the $\Nrf$ layers in Fig.~\ref{fig:fact_graph} is called as a single forward pass in DCS-AMP. A backward pass starts from the $\Nrf ^{\mathrm{th}}$ layer and performs message flows till the first layer is reached. After multiple forward and backward passes, DCS-AMP is expected to converge. At this point, the marginal posterior of $\mathsf{s}_{k}[n]$ for the channel measurements $\{\mathbf{y}_{k}\}^{\Nrf}_{k=1}$ is estimated by multiplying all the messages, i.e., probability densities, received by $\mathsf{s}_{k}[n]$. The mean of the estimated marginal posteriors gives the minimum mean-squared estimate of the sparse masked beamspace subchannels. 
\par In practice, the prior parameters $\epsilon$, $\kappa$, $\xi$, $\zeta$ and $\rho$ are unkown apriori. To overcome this issue, DCS-AMP embeds the message passing algorithm within an Expectation-Maximization (EM) block that learns the parameters online \cite{DCSAMP}. For better subchannel recovery, the identical prior assumption can be waived to learn different set of parameters for different groups of subchannel locations. Inspired by the MRI example in \cite{DCSAMP}, we develop a strategy to group the subchannel locations into two sets that correspond to potentially active and inactive coefficients. Let $\{\hat{\mathbf{S}}_{k,\mathrm{A}}\}^{\Nrf}_{k=1}$ be the subchannel estimates obtained using standard AMP-based CS over each subchannel, i.e., $\hat{\mathbf{S}}_{k,\mathrm{A}}$ is estimated from just $\mathbf{y}_k$. The factor graph corresponding to standard AMP is same as that in Fig.~\ref{fig:fact_graph}, but without the inter-subchannel factors $\{b_k[n]\}_{k,n}$ and $\{q_k[n]\}_{k,n}$. As a result, standard AMP-based CS does not exploit the smooth variation across subchannels and may result in poor subchannel estimates. Our DCS-AMP-based approach uses the subchannel estimates from standard AMP to determine the potentially active and inactive locations. We define an energy matrix corresponding to the subchannels recovered through standard AMP as $\hat{\mathbf{S}}_{\mathrm{E,A}}=\sum_{k=1}^{\Nrf}|\hat{\mathbf{S}}_{k,\mathrm{A}}|^2$. We define $\mathcal{S}_1 \subset \mathcal{I}_N \times \mathcal{I}_N$ as a set that contains the locations of the largest coefficients in $\hat{\mathbf{S}}_{\mathrm{E,A}}$ that contribute to a $\delta_{\mathrm{E}}$ fraction of the total sum in $\hat{\mathbf{S}}_{\mathrm{E,A}}$. The set $\mathcal{S}_2$ contains the locations of the remaining coefficients such that $\mathcal{S}_1 \cup \mathcal{S}_2 = \mathcal{I}_N \times \mathcal{I}_N$. Note that the sets $\mathcal{S}_1$ and $\mathcal{S}_2$ are invariant with the subchannel index $k$. For $\delta_{\mathrm{E}}$ close to $1$, it can be observed that $\mathcal{S}_1$ represents the set of potentially active coefficients. In this paper, we choose $\delta_{\mathrm{E}}=0.9$. The prior parameters in DCS-AMP are learned independently for the masked beamspace locations in $\mathcal{S}_1$ and $\mathcal{S}_2$, using the EM algorithm \cite{DCSAMP}.\footnote{An implementation of DCS-AMP-based subchannel recovery can be found on our page \cite{Proxilink}; the code is based on the MRI example in \cite{DCSAMP}.}
\par We now discuss about the complexity of DCS-AMP with ZC-based subchannel acquisition. In a single pass of DCS-AMP, the complexity of message passing iterations corresponding to each subchannel is determined by matrix-vector multiplications involving $\mathbf{A}_k$ and $\mathbf{A}^{\ast}_k$ \cite{DCSAMP}. It can be observed from \eqref{eq:csmatrixdefn} that the matrix-vector products can be efficiently implemented using the fast Fourier transform, which has a complexity of $\mathcal{O}(N \mathrm{log} N)$. Let $T$ be the number of forward and backward passes that are carried out in DCS-AMP. In each pass of DCS-AMP, message passing iterations are performed for $\Nrf$ subchannels. As a result, the overall complexity of DCS-AMP is $\mathcal{O}( T \Nrf N \mathrm{log} N)$. 
\subsection{Beam alignment and achievable rate} \label{sec:ratecomp}
We now describe how to estimate subchannels using the estimates obtained with DCS-AMP. Let $\{\hat{\mathbf{S}}_{k,\mathrm{D}}\}^{\Nrf}_{k=1}$ be the masked beamspace estimates obtained using DCS-AMP. For a ZC-based spectral mask, the beamspace subchannel estimates $\{\hat{\mathbf{X}}_{k,\mathrm{D}}\}^{\Nrf}_{k=1}$ are found by inverting the spectral mask as $\hat{\mathbf{X}}_{k,\mathrm{D}}=\boldsymbol{\Lambda}^{-1}_{\mathbf{z}}\hat{\mathbf{S}}_{k,\mathrm{D}}\boldsymbol{\Lambda}^{-1}_{\mathbf{z}}$ \cite{swiftlink}. The $k^{\mathrm{th}}$ subchannel estimate derived using DCS-AMP is then $\hat{\mathbf{H}}_{k,\mathrm{D}}=\mathbf{U}_N \hat{\mathbf{X}}_{k,\mathrm{D}} \mathbf{U}_N$. 
\begin{figure}[h!]
\centering
\vspace{-7mm}
\includegraphics[width=0.6 \linewidth]{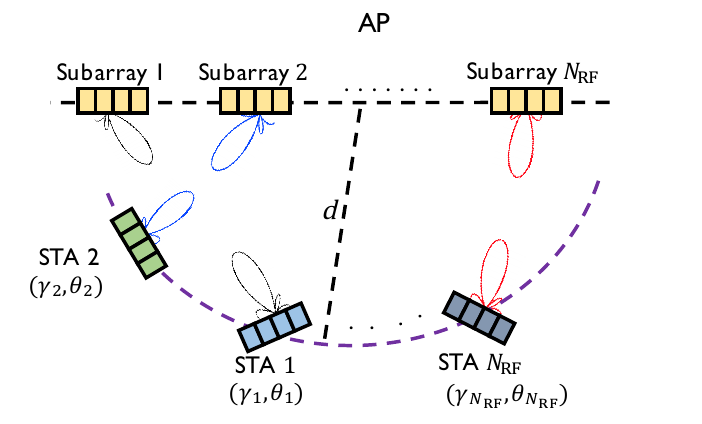}
  \caption{Directional transmission from STAs to the AP in a short range communication system. Each subarray at the AP is associated with a single STA. The inter stream interference that arises after beamforming is mitigated in the digital baseband at the AP.}
  \label{fig:top_view_proxilink}
\end{figure}
\par To evaluate the performance of our DCS-AMP based approach, we consider a setting with $\Nrf$ STAs placed at distinct coordinates around the AP \cite{MMSE}. For simplicity, we assume that all the $\Nrf$ STAs are at the same distance of $d$ from the AP. We use $\mathbf{H}_{k,u} \in \mathbb{C}^{N \times N}$ to denote the subchannel matrix between the $u^{\mathrm{th}}$ STA and the $k^{\mathrm{th}}$ subarray at the AP. Under the perfect synchronization assumption, the $\Nrf$ STAs perform uplink pilot transmissions one at a time for channel estimation. We use $\hat{\mathbf{H}}_{k,u,\mathrm{D}}$ to denote the subchannel estimate corresponding to $\mathbf{H}_{k,u}$, that is obtained with DCS-AMP. At the end of the channel estimation phase, the AP allocates a single RF chain to a STA for data transmission. Without loss of generality, we assume that the $k^{\mathrm{th}}$ RF chain at the AP is allocated to the $k^{\mathrm{th}}$ STA. A graphical illustration of the simulation setting is shown in Fig.~\ref{fig:top_view_proxilink}, where directional uplink data transmission is perfomed between a STA and a single associated RF chain at the AP.
\par Now, we explain how the directional beams at the AP and the STAs are determined from the estimated channel. For DCS-AMP-based channel estimation, the beamforming vectors at the $k^{\mathrm{th}}$ STA and the $k^{\mathrm{th}}$ subarray of the AP are computed from $\hat{\mathbf{H}}_{k,k,\mathrm{D}}$. We define $\mathbf{f}_{k,\mathrm{D}}$ as the $q$-bit phase quantized version of the top right singular vector of $\hat{\mathbf{H}}_{k,k,\mathrm{D}}$, i.e., the singular vector that corresponds to the maximum singular value of $\hat{\mathbf{H}}_{k,k,\mathrm{D}}$ \cite{ZCglobecom}. Similarly, $\mathbf{w}_{k,\mathrm{D}}$ is defined as the conjugate of the $q$-bit phase quantized version of the top left singular vector of $\hat{\mathbf{H}}_{k,k,\mathrm{D}}$. As the beamformers are computed at the AP, the AP must signal the quantized phases to be used at the STAs using a control channel. Beamforming using a predefined codebook like the DFT can help reduce the signalling overhead. In this paper, however, we consider beamforming based on the quantized SVD-based approach to study the best case performance of DCSAMP. During uplink transmission, the STAs simultaneously transmit their data streams to the AP using $\{\mathbf{f}_{k,\mathrm{D}}\}^{\Nrf}_{k=1}$. We use $\mathbf{H}_{\mathrm{UL,D}}\in \mathbb{C}^{\Nrf \times \Nrf}$ to denote the effective multi-user uplink channel seen after analog combining with $\{\mathbf{w}_{k,\mathrm{D}}\}^{\Nrf}_{k=1}$. The $(i,j)^{\mathrm{th}}$ entry of $\mathbf{H}_{\mathrm{UL,D}}$ is given by $\mathbf{H}_{\mathrm{UL,D}}(i,j)=\mathbf{w}^T_{i,\mathrm{D}} \mathbf{H}_{i,j} \mathbf{f}_{j,\mathrm{D}}$. In the data transmission phase, the received signal at the $k^{\mathrm{th}}$ AP subarray comprises of the signal from the $k^{\mathrm{th}}$ STA in addition to interference from the remaining $\Nrf-1$ STAs. The off-diagonal terms in $\mathbf{H}_{\mathrm{UL,D}}$ indicate the amount of interference after analog beamforming using the DCS-AMP-based channel estimates. The interference can be mitigated by appropriate combining in the digital baseband of the AP.
\par In this paper, the classical MMSE beamformer is used at the AP to mitigate interference from other STAs. We assume that all the $\Nrf$ STAs transmit with the same amount of power. In such case, it is reasonable to assume that the received SNR, ignoring interference, is the same at each subarray. For DCS-AMP-based channel recovery, the signal-to-interference-plus-noise ratio (SINR) of the received signal from the $k^{\mathrm{th}}$ STA, after MMSE beamforming, is given by \cite{MMSE}
\begin{equation}
\label{eq:MMSESINR}
\rho_{k,\mathrm{D}}=\frac{\mathrm{SNR}}{[(\mathbf{H}^{\ast}_{\mathrm{UL,D}}\mathbf{H}_{\mathrm{UL,D}}+\mathbf{I}/\mathrm{SNR})^{-1}]_{k,k}}-1.
\end{equation}
The achievable rate corresponding to the $k^{\mathrm{th}}$ user is defined as $R_{k,\mathrm{D}}=\mathrm{log}_2(1+\rho_{k,\mathrm{D}})$. 
\section{Short range link configuration through geometry-aided message passing}  
In typical short range settings, information about the range of the transceiver distance, i.e., $d$ in Fig.~\ref{fig:architect}, may be available. For example, it is reasonable to assume that $d \in [30\, \mathrm{cm}, 130\, \mathrm{cm}]$ in on-body wearable communication settings. In this section, we show that the array geometry of the transceiver together with the distribution on $d$, determine the correlation between the subchannels. The DCS-AMP-based approach, however, does not explicitly incorporate such geometry-based correlation. We construct factors to model such correlation and develop a parametric message passing algorithm that solves for the local AoAs.
\par To explain the idea underlying our method, we consider a single STA setting shown in Fig.~\ref{fig:architect}. Beam alignment with the proposed approach is performed in two stages. In the first stage, pilot transmission is performed in the uplink by using a fixed transmit beam training vector at the STA. During this stage, the subarrays at the AP use different beam training vectors to acquire channel measurements for local AoA estimation. In the second stage, one of the subarrays at the AP performs directional pilot transmission using the local AoA estimated in the first stage. The STA uses different beam training vectors for AoA estimation in the second stage. Finally, the estimated AoA is used for beam alignment at the STA.
 \subsection{Geometry-aware local AoA estimation at the AP} \label{sec:geo_local_AoA}
The proposed geometry-aided message passing algorithm models the local AoAs, i.e., $\{\omega_k\}^{\Nrf}_{k=1}$, as realizations of random variables $\{ \Omega_k \}^{\Nrf}_{k=1}$. Our message passing algorithm uses information from the channel measurements through angle-based likelihoods, and also models the correlation among the angles. 
\subsubsection{Angle-based likelihoods from channel measurements} \label{sec:AoA_likeli}
Channel measurements for local AoA estimation are obtained by using different beam training vectors at the AP. The STA, however, uses a fixed ZC sequence $\mathbf{z}$ as the beam training vector throughout the local AoA estimation process at the AP. The channel response seen by the $k^{\mathrm{th}}$ subarray of the AP is then $\mathbf{h}_{\mathrm{AP},k}=\mathbf{H}_{k}\mathbf{z}$. As $\mathbf{H}_{k} = \alpha_k \mathbf{a}_N(\omega_k)\mathbf{a}^T_N(\phi_k)$ under the far field assumption for the  subchannel, we can write $\mathbf{h}_{\mathrm{AP},k}=\alpha_{\mathrm{AP},k}\mathbf{a}_N(\omega_k)$, where $\alpha_{\mathrm{AP},k}=\alpha_k \mathbf{a}^T_N(\phi_k) \mathbf{z}$ is an unknown constant. The channel measurement acquired when the $k^{\mathrm{th}}$ subarray applies a beam training vector $\mathbf{w}_k[m]$ can be expressed as 
\begin{equation}
\label{eq:meas_local_AP_1}
y_{\mathrm{AP},k}[m]=\alpha_{\mathrm{AP},k}\mathbf{w}^T_k[m]\mathbf{a}_N(\omega_k) +v_k[m].
\end{equation}
The channel measurement in \eqref{eq:meas_local_AP_1} is a noisy projection of $\mathbf{a}_N(\omega_k)$, up to an unknown scaling. The unknown scaling $\alpha_{\mathrm{AP},k}$ in $y_{\mathrm{AP},k}[m]$ is invariant with $m$, due to the use of a fixed ZC sequence at the STA array.
\par We describe the set of beam training vectors used at the AP to obtain channel measurements. We define $M_{\mathrm{AP}}$ as the number of channel measurements acquired by an AP subarray for local AoA estimation. The first $M_{\mathrm{AP}}-1$ projections are acquired by applying distinct circulant shifts $\{r_k[m]\}^{M_{\mathrm{AP}}-1}_{m=1}$ of the ZC sequence $\mathbf{z}$ to the $k^{\mathrm{th}}$ subarray of the AP, i.e., $\mathbf{w}_k[m]=\mathbf{J}_{r_k[m]}\mathbf{z}$. Then, the $M_{\mathrm{AP}}^{\mathrm{th}}$ channel measurement is acquired using $\mathbf{w}_k[M_{\mathrm{AP}}]=\mathbf{w}_k[1]\odot (1,-1,-1, \cdots -1)^T$. The $M_{\mathrm{AP}}^{\mathrm{th}}$ beam training vector is chosen differently to estimate the unknown complex gain $\alpha_{\mathrm{AP},k}$ in \eqref{eq:meas_local_AP_1}. With $\tilde{w}_k$ defined as the first entry of $\mathbf{w}_{k}[1]$, it can be observed from \eqref{eq:meas_local_AP_1} that $y_{\mathrm{AP},k}[1]+y_{\mathrm{AP},k}[M_{\mathrm{AP}}]$ is a noisy version of $2 \alpha_{\mathrm{AP},k} \tilde{w}_k$. The unknown gain $\alpha_{\mathrm{AP},k}$ is estimated as $\hat{\alpha}_{\mathrm{AP},k}=(y_{\mathrm{AP},k}[1]+y_{\mathrm{AP},k}[M_{\mathrm{AP}}])/2 \tilde{w}_k$. We define a projection matrix $\boldsymbol{\Psi}_k \in \mathbb{C}^{M_{\mathrm{AP}} \times N}$ such that $\boldsymbol{\Psi}_k (m,:)=\mathbf{w}^T_k[m]$. The vector of channel measurements in \eqref{eq:meas_local_AP_1} is then $\mathbf{y}_{\mathrm{AP},k}=\alpha_{\mathrm{AP},k} \boldsymbol{\Psi}_k \mathbf{a}_N(\omega_k) +\mathbf{v}_k$. The gain compensated channel measurements are defined as $\tilde{\mathbf{y}}_{\mathrm{AP},k}=\mathbf{y}_{\mathrm{AP},k}/\hat{\alpha}_{\mathrm{AP},k}$. We ignore the error in estimating $\alpha_{\mathrm{AP},k}$ to conclude that $\tilde{\mathbf{y}}_{\mathrm{AP},k}$ is a realization of the random variable $\mathcal{N}_c(\boldsymbol{\Psi}_k \mathbf{a}_N(\omega_k), \sigma^2 \mathbf{I} / | \hat{\alpha}_{\mathrm{AP},k}|^2)$. 
\par The likelihood of the local AoA $\omega_k$ is computed from the gain compensated channel measurements as 
\begin{equation}
\label{eq:likelihood_1}
p(\omega_k)=\mathrm{exp}\left(-{| \hat{\alpha}_{\mathrm{AP},k}|^2 \Vert \tilde{\mathbf{y}}_{\mathrm{AP},k} - \boldsymbol{\Psi}_k \mathbf{a}_N(\omega_k) \Vert_2^{2}}/{\sigma^2}\right).
\end{equation}
A reasonable approach to estimate the local AoAs $\{\omega_k\}_{k=1}^{\Nrf}$ is to maximize the likelihood functions $\{p(\omega_k)\}_{k=1}^{\Nrf}$ independently. We define the maximum likelihood (ML)-based local AoA estimate as $\hat{\omega}_{k,\mathrm{ML}}=\mathrm{arg\,max}\,p(\omega_k)$. The ML-based approach, however, does not exploit the joint dependency among the local AoAs. As a result, it is possible that such an approach can result in local AoA estimates that are inconsistent with the geometry of the problem. For example, it can be observed from Fig.~\ref{fig:architect} that the  combination $\omega_1=\pi/4$ and $\omega_2=-\pi/4$ is infeasible according to the geometry. Therefore, it is important to account for the dependencies among the local AoAs, that arise from the geometry, in the estimation technique. 
\subsubsection{Geometry-based dependencies for local AoA estimation}
In this section, we model the dependencies among the local AoAs by constructing conditional probability densities of $\{\Omega_k\}_{k=1}^{\Nrf}$. Without loss of generality, we consider the local AoAs $\omega_1$ and $\omega_2$ to explain our construction. The distance between the midpoints of the STA and the first AP subarray is defined as $d_1$. We use $\ell_k$ to denote the  distance between the midpoint of the $k^{\mathrm{th}}$ AP subarray and the midpoint of the AP array. For a particular $d$, the distance $d_{1}$ in Fig.~\ref{fig:architect} can be obtained from
\begin{equation}
\label{eq:d1_solve}
(d_{1}\mathrm{cos}\,\omega_{1})^{2}+(d_{1}\mathrm{sin}\,\omega_{1}+\ell_{1})^{2}=d^{2}
\end{equation} 
and, subsequently, the local AoA $\omega_2$ can be expressed as 
\begin{equation}
\label{eq:omega2solve}
\omega_2=\mathrm{tan}^{-1}\left(\frac{\ell_1-\ell_2+d_{1}\mathrm{sin}\,\omega_{1}}{d_{1}\mathrm{cos}\,\omega_{1}}\right).
\end{equation}
The relation between $\omega_1$ and $\omega_2$ is determined by $\omega_2= \mathcal{G}(\omega_1,d)$, where $\mathcal{G}$ is a function defined by \eqref{eq:d1_solve} and \eqref{eq:omega2solve}. It can be observed from \eqref{eq:d1_solve} and \eqref{eq:omega2solve} that there is a unique $\omega_2$ for a particular $\omega_1$ and $d$. 
\par We construct factors for a realistic scenario in which the distance $d$ is unknown. Under the assumption that $d\in [d_{\mathrm{min}},d_{\mathrm{max}}]$, the range of $\omega_2$ for a given $\omega_1$ can be determined using \eqref{eq:d1_solve} and \eqref{eq:omega2solve}. As the local AoAs are considered as realizations of random variables, we capture the local AoA dependencies using conditional distributions. We use $p_\mathrm{g}(\omega_2|\omega_1)$ to denote the distribution of $\Omega_2$ conditioned on $\Omega_1= \omega_1$. We assume that $d$ is uniformly distributed in $[d_{\mathrm{min}},d_{\mathrm{max}}]$ to write 
\begin{equation}
\label{eq:geo_fact}
p_\mathrm{g}(\omega_2|\omega_1)=\frac{1}{D}\int_{d_{\mathrm{min}}}^{d_{\mathrm{max}}}  \!\!\!\! \delta(\omega_2-\mathcal{G}(\omega_{1},r))dr,
\end{equation}  
where $D=d_{\mathrm{max}}-d_{\mathrm{min}}$, and the Dirac-delta function in \eqref{eq:geo_fact} indicates $\omega_2=\mathcal{G}(\omega_1,r)$. The procedure to determine $p_\mathrm{g}(\omega_2|\omega_1)$, can be used to compute other conditional distributions, i.e., $\{p_{\mathrm{g}}(\omega_k|\omega_n)\}_{k,n}$. It is important to note that the conditional distributions only depend on the geometry of the AP array and prior information about the transceiver distance. Therefore, the conditional distributions $\{p_{\mathrm{g}}(\omega_k|\omega_n)\}_{k,n}$ can be computed offline, i.e., before acquiring channel measurements.
\subsubsection{Message passing algorithm at the AP}
\begin{figure}[h!]
   \centering
   \includegraphics[width=0.55 \textwidth]{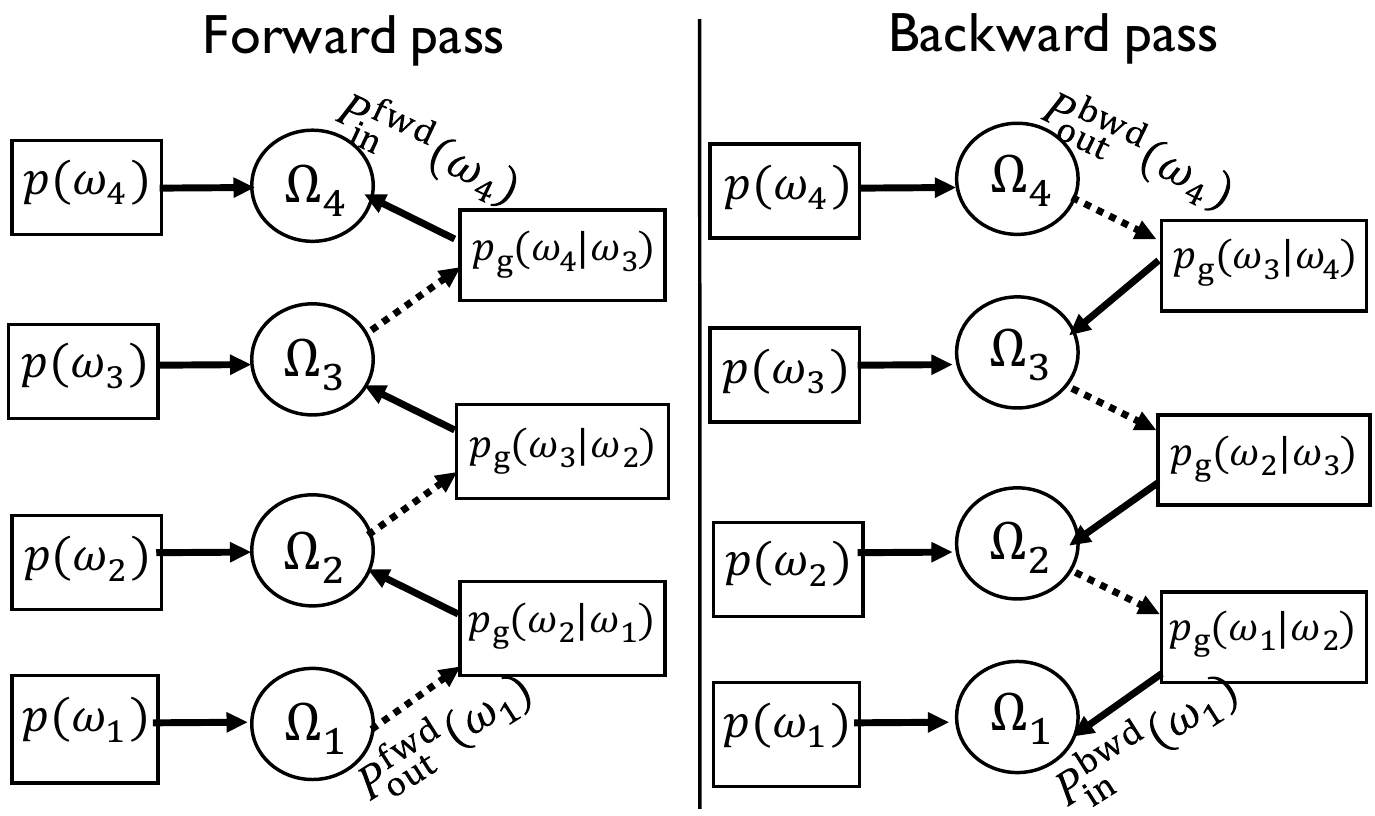}
   \caption{Factor graphs in geometry-aided message passing for local AoA estimation at the AP. Here, $\Nrf=4$. The random variable $\Omega_k$ models a realization of the local AoA $\omega_k$. Message flows are performed between the local AoA likelihoods, i.e., $\{p(\omega_k)\}^{\Nrf}_{k=1}$, and the geometry-based factors, i.e., $p_{\mathrm{g}}(\cdot)$.}
   \label{fig:fact_geo}
\end{figure}

We now describe the factor graph in Fig.~\ref{fig:fact_geo} for our geometry-aided message passing algorithm. The factor graph consists of circular nodes, called as variable nodes, that represent the random variables $\{\Omega_k\}_{k=1}^{\Nrf}$. The rectangular nodes in Fig.~\ref{fig:fact_geo}, called as factors, contain the likelihoods or the conditional local AoA probabilities. Our algorithm performs message flows between the variable nodes and the factors to estimate the local AoAs. The message flows include a single forward pass and a single backward pass, and have a similar structure to the flows in DCS-AMP. 
\par Now, we explain how messages are constructed in the forward pass. In the first step of the forward pass, the local AoA likelihood $p(\omega_1)$ is sent to node $\Omega_1$. The message received by node $\Omega_1$ can be thought of a probability distribution of $\Omega_1$ that is believed by the factor based on the channel measurements. The node $\Omega_1$ then forwards $P^{\mathrm{fwd}}_{\mathrm{out}}(\omega_{1})=p(\omega_1)$ to the geometry factor that contains $p_{\mathrm{g}}(\omega_2|\omega_1)$, so that the belief about $\Omega_2$ can be computed. 
The message $P^{\mathrm{fwd}}_{\mathrm{in}}(\omega_{2})$, sent from the geometry factor to node $\Omega_2$, is defined as
\begin{equation}
\label{eq:pfwd_in_theta2}
P^{\mathrm{fwd}}_{\mathrm{in}}(\omega_{2})=\int_{-\pi/2}^{\pi/2}P^{\mathrm{fwd}}_{\mathrm{out}}(\omega_1)p_{\mathrm{g}}(\omega_{2}|\omega_1)d\omega_{1}.
\end{equation}
It can be observed from \eqref{eq:pfwd_in_theta2} that $P^{\mathrm{fwd}}_{\mathrm{in}}(\omega_{2})$ is $p_{\mathrm{g}}(\omega_{2}|\omega_{1})$ marginalized over $\omega_1$ using the belief from node $\Omega_1$, i.e., $P^{\mathrm{fwd}}_{\mathrm{out}}(\omega_{1})$.  The message $P^{\mathrm{fwd}}_{\mathrm{in}}(\omega_{2})$ can be interpreted as a side information about $\Omega_2$ as it does not depend on the measurements acquired by the second subarray of the AP. This side information is generated using the channel measurements in the first subarray of the AP, and geometry-based dependency between $\omega_1$ and $\omega_2$. The node $\Omega_2$ combines information from the channel measurements, i.e., $p(\omega_2)$, and the side information $P^{\mathrm{fwd}}_{\mathrm{in}}(\omega_{2})$ by constructing 
\begin{equation}
\label{eq:pfwd_out_theta2}
P^{\mathrm{fwd}}_{\mathrm{out}}(\omega_2)=p(\omega_2)P^{\mathrm{fwd}}_{\mathrm{in}}(\omega_{2}).
\end{equation} 
The message in \eqref{eq:pfwd_out_theta2} is then sent to the factor containing $p_{\mathrm{g}}(\omega_3|\omega_2)$, which computes the belief about $\Omega_3$ using $p_{\mathrm{g}}(\omega_3|\omega_2)$ and $P^{\mathrm{fwd}}_{\mathrm{out}}(\omega_2)$. The message sent by the factor, i.e., $P^{\mathrm{fwd}}_{\mathrm{in}}(\omega_{3})$, is computed using an expression similar to \eqref{eq:pfwd_in_theta2}. The process of message flows is executed until the last node $\Omega_{\Nrf}$ is reached. The messages in \eqref{eq:pfwd_in_theta2} and \eqref{eq:pfwd_out_theta2} can be generalized, by setting $\omega_1$ to $\omega_{k-1}$ and $\omega_{2}$ to $\omega_{k}$, to obtain recursive equations. The forward pass computes the message inflow $\{P^{\mathrm{fwd}}_{\mathrm{in}}(\omega_{k})\}_{k=2}^{\Nrf}$ using the generalized recursive equations.  
\par The forward pass performs message flows in a particular direction, i.e., increasing $k$. As a result, it does not exploit information from subarrays $k,k+1,\cdots,\Nrf$, to generate side information about $\Omega_{k-1}$. Now, we describe the second component of our message passing algorithm, i.e., the backward pass. Message flows in the backward pass 
begin at node $\Omega_{\Nrf}$, and are performed until node $\Omega_1$ is reached. In the first step of the backward pass, the 
likelihood derived from the channel measurements, i.e., $P^{\mathrm{bwd}}_{\mathrm{out}}(\omega_{\Nrf})=p(\omega_{\Nrf})$, flows into the geometry factor containing $p_{\mathrm{g}}(\omega_{\Nrf-1}|\omega_{\Nrf})$. Then, the geometry factor 
computes a belief about $\Omega_{\Nrf-1}$ using $P^{\mathrm{bwd}}_{\mathrm{out}}(\omega_{\Nrf})$ and $p_{\mathrm{g}}(\omega_{\Nrf-1}|\omega_{\Nrf})$. The belief about $\Omega_{\Nrf-1}$ is denoted by $P^{\mathrm{bwd}}_{\mathrm{in}}(\omega_{\Nrf-1})$. The messages in the backward pass are computed using the recursive equations   
\begin{align}
\label{eq:pbwd_in_thetak}
P^{\mathrm{bwd}}_{\mathrm{in}}(\omega_{k-1})&=\int_{-\pi/2}^{\pi/2}P^{\mathrm{bwd}}_{\mathrm{out}}(\omega_{k})g(\omega_{k-1}|\omega_{k})d\omega_{k},\\
\label{eq:pbwd_out_prod_k}
P^{\mathrm{bwd}}_{\mathrm{out}}(\omega_{k})&=p(\omega_k)P^{\mathrm{bwd}}_{\mathrm{in}}(\omega_{k}).
\end{align}
It can be observed from Fig.~\ref{fig:fact_geo} that the nodes $\Omega_1$ and $\Omega_{\Nrf}$ do not receive any side information in the forward and backward passes. To validate the recursive equations, we define $P^{\mathrm{fwd}}_{\mathrm{in}}(\omega_{1})=\mathcal{U}(\omega_1)$ and $P^{\mathrm{bwd}}_{\mathrm{in}}(\omega_{\Nrf})=\mathcal{U}(\omega_{\Nrf})$, where $\mathcal{U}(\omega)$ denotes a uniform distribution over $\omega$.   
\par The final component of our message passing algorithm combines information from the forward and backward passes in a manner that is similar to DCS-AMP. It can be observed that node $\Omega_k$ receives side information in the form of $P^{\mathrm{fwd}}_{\mathrm{in}}(\omega_{k})$ and $P^{\mathrm{bwd}}_{\mathrm{in}}(\omega_{k})$ in the forward and backward passes. Furthermore, it also has access to the likelihood $p(\omega_k)$. These three sources of information about $\Omega_k$ are combined by defining a new belief 
\begin{equation}
\label{eq:combprob}
p_{_{\mathrm{GMP}}}(\omega_k)=p(\omega_k)P^{\mathrm{fwd}}_{\mathrm{in}}(\omega_k)P^{\mathrm{bwd}}_{\mathrm{in}}(\omega_k).
\end{equation}
The local AoA estimate, using our geometry-aided message passing algorithm, is defined as $\hat{\omega}_k=\mathrm{arg\,max}\,p_{_{\mathrm{GMP}}}(\omega_k)$. 
\par Our implementation considers a discrete support for local AoAs to compute the message flows in the forward and backward passes. We assume an angular resolution of $\delta_{\omega} \pi$, for each of the local AoAs, to compute the integrals in \eqref{eq:geo_fact}, \eqref{eq:pfwd_in_theta2}, and \eqref{eq:pbwd_in_thetak} using a discrete sum. The complexity of our algorithm is determined by the integration step in message passing. For example, it can be observed that the discrete version of \eqref{eq:pbwd_in_thetak} requires summing up $\mathcal{O}(1/\delta_{\omega})$ terms for each candidate $\omega_{k-1}$. As there are $\mathcal{O}(1/\delta_{\omega})$ candidates for $\omega_{k-1}$, the complexity of the integration step is $\mathcal{O}(1/\delta^2_{\omega})$. As $\Nrf-1$ such integrations are performed in each of the forward and backward passes, the overall complexity of our algorithm is $\mathcal{O}(\Nrf/\delta^2_{\omega})$. We would like to highlight that the proposed technique requires a single forward and backward pass, unlike DCS-AMP that performs multiple passes. The factor graph in both these techniques only model the correlation between adjacent subchannels. For example, the proposed technique does not include factors like $p_{\mathrm{g}}(\omega_k|\omega_\ell)$ for $|k-\ell|>1$. Incorporating such factors can result in short cycles that may not be desirable from a message passing perspective \cite{cycles_avoid}.
\subsection{Link configuration in a multi-user setting} \label{sec:rate_geo_mp}
\par We explain how our geometry-aided message passing algorithm is applied in a multi-user setting in Fig.~~\ref{fig:top_view_proxilink}. The signal structure used for our algorithm is similar to the one in the IEEE 802.11ad standard \cite{11ad}. First, the STAs transmit pilots so that the AP can estimate the local AoAs using the proposed algorithm. A single STA is assumed to be active at any point of time during the uplink pilot transmission phase. Such an assumption can be relaxed by using orthogonal pilot sequences, in the time or frequency dimensions, at different STAs. Coordination among the STAs for uplink signalling can be achieved with scheduling using the control channel. We use $\omega_{k,u}$ to denote the local AoA at the $k^{\mathrm{th}}$ AP subarray corresponding to the $u^{\mathrm{th}}$ STA. For the geometry-aided message passing algorithm, the local AoA estimate corresponding to $\omega_{k,u}$ is denoted by $\hat{\omega}_{k,u,{\mathrm{GMP}}}$. The AP dedicates its $k^{\mathrm{th}}$ subarray to the $k^{\mathrm{th}}$ STA for downlink signalling and data communication. To achieve directional beamforming at the AP, a $q$-bit phase quantized version of $\overline{\mathbf{a}_N}(\hat{\omega}_{k,k,{\mathrm{GMP}}})$ is used as the beamforming vector at the $k^{\mathrm{th}}$ subarray of the AP. The subarrays at the AP perform downlink pilot transmissions using these directional beams so that the AoA at the STAs can be determined. 
\par In the downlink pilot transmission phase, each STA acquires $M_{\mathrm{STA}}$ channel measurements by applying distinct random circulant shifts of $\mathbf{z}$ to its phased array. We use $\phi_{k,k}$ to denote the AoA at the $k^{\mathrm{th}}$ STA that is associated with the $k^{\mathrm{th}}$ subarray at the AP. Similar to the procedure in Sec. \ref{sec:AoA_likeli}, the $k^{\mathrm{th}}$ STA computes the likelihood function $p(\phi_{k,k})$ with the received channel measurements. The AoA estimate at the $k^{\mathrm{th}}$ STA is given by $\hat{\phi}_{k,k,\mathrm{ML}}=\mathrm{arg\,max}\,p(\phi_{k,k})$. Using the estimated local AoA, the $k^{\mathrm{th}}$ STA performs beam alignment by applying a $q$-bit phase quantized version of $\overline{\mathbf{a}_N}(\hat{\phi}_{k,k,\mathrm{ML}})$ to its phased array. It is important to note that the AP must feedback information about the best beamformers to the STAs in the DCS-AMP-based approach. The proposed method, however, does not require any such feedback as the beamformers are computed at the receiving end. A graphical illustration of multi-user data transmission with the estimated beamformers at the AP and the STAs is shown in Fig.~~\ref{fig:top_view_proxilink}. We define the $\mathbf{H}_{\mathrm{UL,GMP}} \in \mathbb{C}^{\Nrf \times \Nrf}$ as the multi-user uplink channel seen after configuring the phased arrays at the AP and the STA. The $(i,j)^{\mathrm{th}}$ entry of $\mathbf{H}_{\mathrm{UL,GMP}}$ is given by $\mathbf{H}_{\mathrm{UL,GMP}}(k,u)=\mathbf{a}^{\ast}_N(\hat{\omega}_{k,k,{\mathrm{GMP}}}) \mathbf{H}_{k,u}\overline{\mathbf{a}_N}(\hat{\phi}_{u,u,\mathrm{ML}})$. Finally, an MMSE beamformer is used at the AP to cancel the interference that arises in the multi-user setting. The SINR and the achievable rate after MMSE beamforming are computed by using $\mathbf{H}_{\mathrm{UL,GMP}}$ in \eqref{eq:MMSESINR}.
\section{Simulations}\label{sec:simulations}
\par In this section, we evaluate the proposed techniques in a realistic communication setting that also has non-LoS components. We consider a hardware architecture in Fig.~\ref{fig:architect}, where the AP is equipped with $\Nrf=4$ subarrays. We assume that each subarray at the AP has $N=16$ antennas. The resolution of the phase shifters at the AP and the STA is set as $q=2\, \mathrm{bits}$. The carrier frequency of the mmWave system is set to $60\, \mathrm{GHz}$, that corresponds to a wavelength of $\lambda= 5\, \mathrm{mm}$. It can be observed that each subarray is of length $3.75\, \mathrm{cm}$. We assume that the length of the antenna array at the AP is $L_{\mathrm{AP}}=20\, \mathrm{cm}$. It can be shown that the spacing between consecutive subarrays at the AP is about $1\, \mathrm{cm}$. The AP considered in this setting is practical for wearable applications as its length, i.e., $L_{\mathrm{AP}}$, is comparable to the typical length of augmented reality or virtual reality headsets. 
\begin{figure}[h!]
\vspace{-3mm}
\centering
\includegraphics[width=0.5 \textwidth]{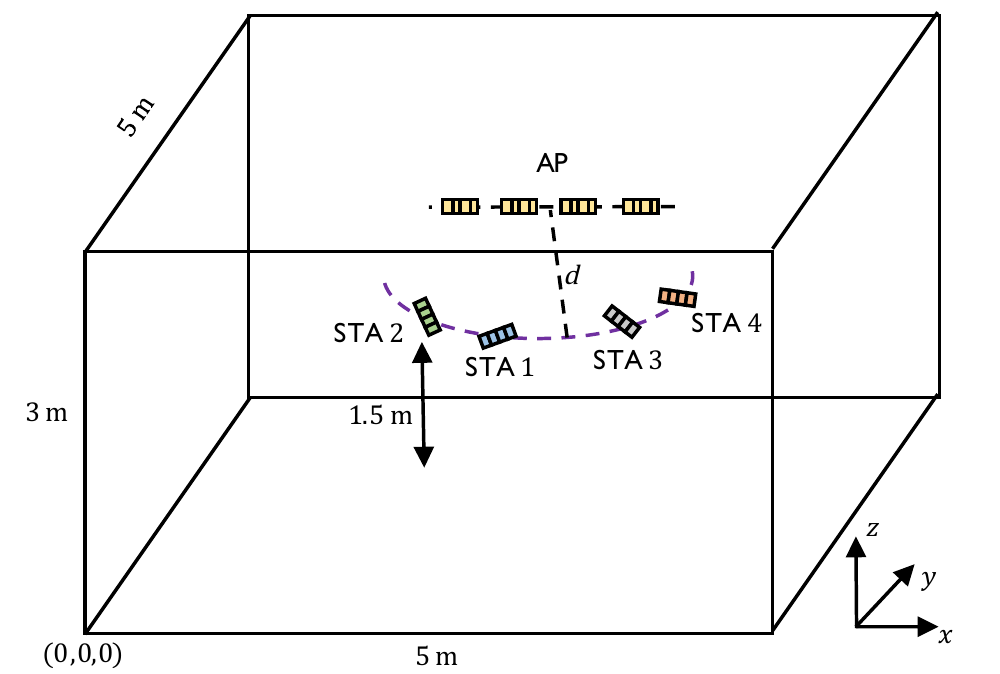}
  \caption{A 3D view of the channel environment considered in our simulations. The AP and the STAs lie on the same horizontal plane of height $1.5 \, \mathrm{m}$. All the STAs are assumed to be at a distance of $d$ from the AP. The top view of this 3D setting is shown in Fig.~\ref{fig:top_view_proxilink}. The channel model in our simulations also includes reflections due to the walls. }
  \label{fig:3D_view_proxilink}
\end{figure}
\par We consider $4$ STAs, where each STA is equipped with a phased array of $N=16$ elements. The antenna array at the STA occupies a length of $L_{\mathrm{STA}}=4\,\mathrm{cm}$, and can be realized in wearables like smart watches and fitness trackers. We assume that AP and the STA arrays are coplanar, and operate at the same polarization. For the $u^{\mathrm{th}}$ STA, the angles corresponding to $\gamma$ and $\theta$ in Fig.~\ref{fig:architect} are labeled as $\gamma_u$ and $\theta_u$. The location of the $u^{\mathrm{th}}$ STA relative to the AP is determined by the triplet $(d, \gamma_u, \theta_u)$. We assume that $|\gamma_u|\leq 75^{\mathrm{o}} $ for any $u$. The angles $\{ \gamma_u \}_{u=1}^{\Nrf}$ are randomly chosen without repetition from the discrete set $\{-75^{\mathrm{o}} +15^{\mathrm{o}}i: i \in \mathcal{I}_{9}\}$. The orientation of the STAs relative to the AP array, i.e., $\{ \theta_u \}_{u=1}^{\Nrf}$, are chosen at random from the discrete set $\{10^{\mathrm{o}}i: i \in \mathcal{I}_{18}\}$. The simulation results we report are for $1000$ random configurations of the STAs chosen according to this distribution; this corresponds to $1000$ different channel realizations. A graphical illustration of the short range mmWave setting is shown in Fig.~\ref{fig:top_view_proxilink} for a particular realization of $\{ ( \gamma_u, \theta_u ) \}_{u=1}^{\Nrf}$. 
\par Now, we describe the indoor short range communication scenario that is considered for our simulations. The AP and the STAs are placed in a room of floor area $5\,\mathrm{m}\, \times 5\,\mathrm{m}$ and a height of $ 3\,\mathrm{m}$, as shown in Fig.~\ref{fig:3D_view_proxilink}. The AP is placed on the sidewall given by $y= 5 \, \mathrm{m}$ at a height of $1.5\, \mathrm{m}$, and is oriented along the $x$-axis. All the $\Nrf$ STAs are placed at a distance of $d=80\, \mathrm{cm}$ from the AP, on the horizontal plane given by $z= 1.5\, \mathrm{m}$. The center of the AP array is at $(2.5\, \mathrm{m},\,5\, \mathrm{m},\,1.5\, \mathrm{m})$. The $N\Nrf \times N$ channel matrix $\mathbf{H}$ is computed using the geometric-ray-based approach in \cite{roomchannelmodel}. Investigating the performance of our algorithm with more realistic channel models, which account for blockage and spatio-temporal channel dynamics \cite{slezak2018empirical}, is an interesting research direction. For each combination of antennas at the AP and the STA, the geometric approach in \cite{roomchannelmodel} considers three kinds of propagation rays. The first kind is due to the direct path between the antennas, and the channel entry that results from such a ray is given in \eqref{eq:shortchan}. The second type of propagation is determined by one bounce reflections in the environment. It is assumed that the floor is carpeted, and reflections from the floor are ignored due to high scattering by carpeted surfaces \cite{carpetrefl}. As a result, there are three one bounce reflections between a pair of antennas, i.e., one due to the ceiling and two due to the sidewalls. The third kind of propagation is determined by two bounce reflections and consists of $4$ rays. These rays arise due to reflections from a pair of sidewalls, and reflections from ceiling and sidewalls. In our simulations, the reflection parameters of the walls and the ceiling were set according to \cite{roomchannelmodel}. For a detailed description on the geometric short range channel model, we refer the reader to \cite{roomchannelmodel}. We observed that the coefficients of the beamspace subchannels corresponding to this indoor channel model exhibit sparsity and smooth variation across the subchannel index. 
\par Channel measurements in the DCS-AMP-based approach and the geometry-aided message passing approach are acquired using circulant shifts of a ZC sequence $\mathbf{z}$. The root of the ZC sequence, i.e., $t$ in \eqref{eq:ZCdefn}, is chosen as $9$ to ensure that $\mathbf{z}$ can be realized in the 2-bit phased array \cite{ZCglobecom}. In the DCS-AMP-based method, the AP acquires $M\Nrf$ ZC-based projections of the channel matrix associated with each STA. Pilot transmissions are only performed in the uplink for link configuration using DCS-AMP. For the geometry-aided message passing appproach, $M_{\mathrm{AP}}$ and $M_{\mathrm{STA}}$ pilot transmissions are perfomed in the uplink and the downlink for each STA. In this paper, we use $M_{\mathrm{AP}}=M/2$ and $M_{\mathrm{STA}}=M/2$ so that the total number of channel measurements corresponding to each STA is same as that in the DCS-AMP-based approach. The resolution in the AoA domain, for the geometry-aided technique, is chosen as $1^{\mathrm{o}}$, i.e., $\delta_{\omega}=1/180$. Furthermore, the geometry factors were computed by assuming $d_{\mathrm{min}}=30\, \mathrm{cm}$ and $d_{\mathrm{max}}=130\, \mathrm{cm}$. The number of forward and backward passes in DCS-AMP is set to $T=12$. The reported results are for several random realizations of random circulant shifts at the AP and the STA, for each channel realization. The performance of the two techniques is evaluated in terms of the achievable rate per STA.  
\par We compare the performance of the proposed link configuration algorithms with three different benchmarks. The first one is based on standard AMP-based CS, that solves for each sparse masked beamspace subchannel independently using the EM-BG-AMP algorithm \cite{EMGAMP}. The same ZC-based CS matrices are used for both DCS-AMP and standard AMP-based techniques. The second benchmark is based on maximum likelihood-based local AoA estimation at the AP. In this approach, the local AoA in a subarray is determined only from the channel measurements acquired in that subarray. The two benchmarks allow us to study the use of inter-subchannel factors in DCS-AMP, and geometry factors in the proposed geometry-aided message passing techniques. Finally, the third benchmark is based on the perfect CSI scenario in which the beamformers are derived from $\mathbf{H}_{k,k}$. The rate achieved with each of the three benchmarks is computed using the procedure in Sec. \ref{sec:ratecomp}.
\begin{figure}[h!]
\centering
\includegraphics[trim=1.5cm 6.25cm 2cm 7.5cm,clip=true,width=0.55 \textwidth]{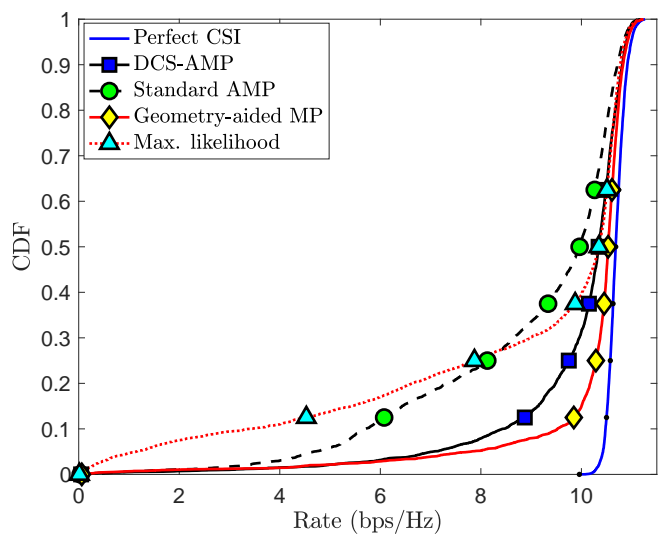}
  \caption{The proposed techniques achieve better rates over standard AMP- and ML-based techniques as they account for the correlation across subchannels using appropriate factors. The results in this plot are for an SNR of $10\, \mathrm{dB}$ and $M=16$ pilot transmissions per STA.}
  \label{fig:CDF_plot_proxilink}
\end{figure}
\par
\begin{figure}[h!]
\vspace{-1mm}
\centering
\includegraphics[trim=2cm 6.75cm 2.5cm 7.5cm,clip=true,width=0.55 \textwidth]{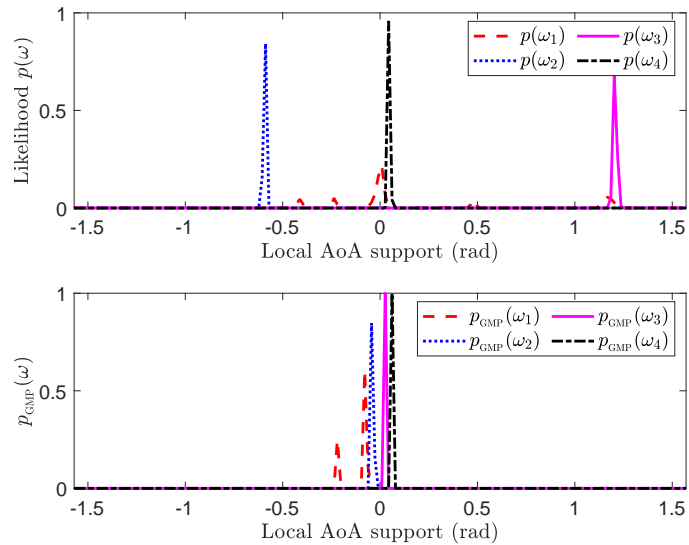}
  \caption{Likelihoods for local AoAs at the AP, and the distributions $\{p_{_{\mathrm{GMP}}}(\omega_k)\}_{k=1}^{\Nrf}$, for a channel realization in which $\gamma=0$ and $\theta=0$. Here, $\mathrm{SNR}=10\, \mathrm{dB}$ and $M_{\mathrm{AP}}=4$. The locations where $\{p_{_{\mathrm{GMP}}}(\omega_k)\}_{k=1}^{\Nrf}$ achieve maximum are close to the true local AoAs, when compared to those that maximize the likelihoods.}
  \label{fig:aoa_supp_plot}
    \vspace{-1mm}
\end{figure}

 As each subchannel has dimensions of $16 \times 16$, it can be observed that beam alignment through conventional MIMO channel estimation requires $M=256$ pilot transmissions from every STA. The proposed techniques can be used to perform beam alignment with fewer channel measurements, i.e., $M<256$. In Fig.~\ref{fig:CDF_plot_proxilink}, we plot the empirical cumulative distribution function (CDF) of the achievable rate obtained with the proposed methods and the benchmarks. The achievable rate considered in our results is a concatenated version of the per-user rates. For $M=16$ pilot transmissions per STA and an SNR of $10 \, \mathrm{dB}$, the CDFs in Fig.~\ref{fig:CDF_plot_proxilink} indicate that the message passing-based techniques can be used to achieve a reasonable rate with sub-Nyquist channel measurements. DCS-AMP achieves a better rate over standard AMP as it exploits the smooth subchannel variation in addition to sparsity. It can be observed from Fig.~\ref{fig:CDF_plot_proxilink} that the proposed geometry-aided message passing algorithm with $M_{\mathrm{AP}}=M_{\mathrm{STA}}=8$ performs better beam alignment when compared to the maximum likelihood-based approach, as it exploits additional local AoA dependencies using the geometry factors. The plot in Fig.~\ref{fig:CDF_plot_proxilink} shows that the proposed message passing method outperforms DCS-AMP for short range link configuration. It is because our algorithm exploits the structure across the subchannels explicity using geometry factors, unlike the DCS-AMP-based approach that models subchannel variation using a Gauss-Markov process. For a particular STA and a channel realization, the likelihoods $\{p(\omega_k)\}_{k=1}^{\Nrf}$ and the distributions $\{p_{_{\mathrm{GMP}}}(\omega_k)\}_{k=1}^{\Nrf}$ are shown in Fig.~\ref{fig:aoa_supp_plot}. It can be noticed from Fig.~\ref{fig:aoa_supp_plot} that geometry-aided message passing reduces the uncertainity in local AoAs when compared to the ML-based approach.
\par
\begin{figure}[h!]
\centering
\includegraphics[trim=1.5cm 6.25cm 2cm 7.5cm,clip=true,width=0.55 \textwidth]{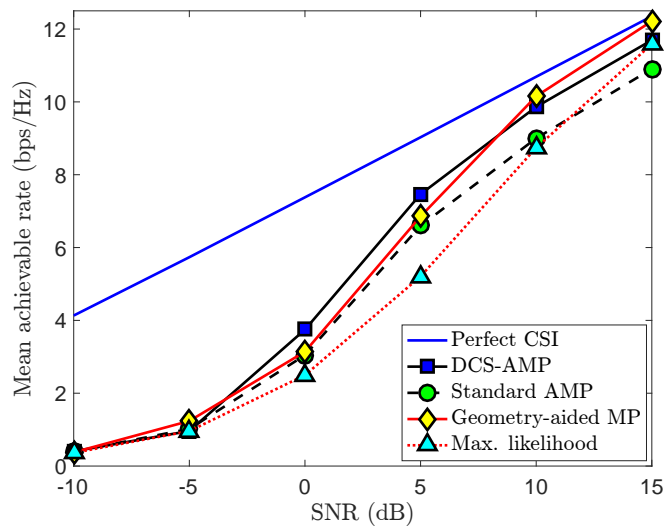}
  \caption{The plot shows the achievable rates as a function of the SNR for $M=16$ pilot transmissions per STA. The rate is computed after MMSE beamforming with the procedures in Sections ~\ref{sec:ratecomp} and~\ref{sec:rate_geo_mp}. For a fair comparison, we use $M_{\mathrm{AP}}=8$ and $M_{\mathrm{STA}}=8$ in the geometry-aided message passing technique. DCS-AMP achieves a higher rate than geometry-aided message passing in the low SNR regime.}
  \label{fig:Rate_vs_SNR}
  \vspace{-2mm}
\end{figure}

 Now, we study the performance of the short range link configuration algorithms as a function of the SNR. The results in Fig.~\ref{fig:Rate_vs_SNR} are for $M=16$. An interesting observation from Fig.~\ref{fig:Rate_vs_SNR} is that DCS-AMP algorithm performs better than the proposed geometry-aided algorithm for SNRs of $0\,\mathrm{dB}$ and $5\,\mathrm{dB}$. At low SNRs, the poor performance of our geometry-based algorithm is mainly due to two reasons. First, the estimation errors in $\alpha_{\mathrm{AP},k}$, at low SNRs, impact the likelihoods in \eqref{eq:likelihood_1} and our message passing algorithm. Second, the proposed geometry-based technique uses a fixed transmit beam training vector at the STA, i.e., $\mathbf{z}$,  for local AoA estimation at the AP. As a result, it is susceptible to nulls in the quasi-omnidirectional beam at the STA, especially at low SNRs. The DCS-AMP-based approach is robust to such nulls due to the use of different shifted ZC sequences $\mathbf{z}$ at the STA, i.e., the location of the nulls vary across different beam  configurations at the STA.
\par
\begin{figure}[h!]
\vspace{-2mm}
\centering
\includegraphics[trim=1.5cm 6.5cm 2cm 7.5cm,clip=true,width=0.55 \textwidth]{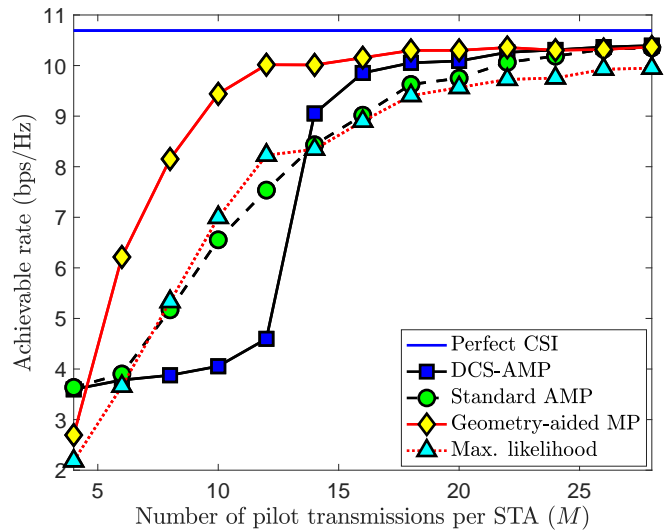}
  \caption{ The achievable rates as a function of $M$, i.e., the number of pilot transmissions, for an SNR of $10\, \mathrm{dB}$. DCS-AMP performs better than standard AMP for a wide range of $M$. For $M <14$, the poor performance of DCS-AMP when compared to standard AMP can be attributed to the smoothening effect.}
  \label{fig:Rate_vs_Meas}
\end{figure}

 We investigate how the achievable rate varies with the number of pilot transmissions using Fig.~\ref{fig:Rate_vs_Meas}. It can be observed from Fig.~\ref{fig:Rate_vs_Meas} that DCS-AMP results in poor performance when compared to standard AMP for $M<14$. The transition effect in Fig.~\ref{fig:Rate_vs_Meas} is best understood by studying the impact of introducing the inter-subchannel factors to the factor graph of standard AMP; such factors ensure that the neighbouring subchannels are not significantly different from each other. Similar to any CS algorithm, the probability that standard AMP fails to recover some subchannels is higher for a lower $M$. When the number of subchannels that cannot be recovered by standard AMP is significantly smaller than $\Nrf$, the inter-subchannel factors in DCS-AMP ``correct'' the poor subchannel estimates obtained with standard AMP, by using information from other subchannels. The correction is likely to be successful if the neighbouring subchannels can be successfully recovered with standard AMP. For a small enough $M$ for which standard AMP fails to recover a significant number of subchannels, introducing the inter-subchannel factors in DCS-AMP can corrupt the successfully recovered subchannels. As a result, DCS-AMP performs poorly when compared to standard AMP for a sufficiently small $M$. 
\par From Fig.~\ref{fig:Rate_vs_Meas}, it can be observed that the proposed geometry-aided message passing algorithm achieves near optimal rates with fewer pilot transmissions, i.e., $M\approx10$. The proposed algorithm has a lower complexity than DCS-AMP due to two reasons. First, our algorithm operates on a lower dimensional representation of the MIMO channel, i.e., using local AoAs, when compared to DCS-AMP that performs optimization over the full MIMO channel. Second, a single forward and a single backward pass is carried out in our algorithm, in comparison to DCS-AMP that uses multiple passes. As seen in Fig.~\ref{fig:Rate_vs_Meas}, the geometry-based approach performs better than the ML-based approach as it exploits the dependencies among the local AoAs. It is important to note, however, that the geometry-aided message passing algorithm assumes a LoS scenario and may perform poorly in channels that have strong non-LoS components. In such settings, DCS-AMP algorithm can perform better than the geometry-based approach as it can also handle the non-LoS components.   
\section{Conclusions and Future work}
Short range mmWave channels exhibit different structural properties when compared to the commonly studied far field channels. For instance, a typical short range LoS MIMO channel can have a rank that is higher than one. The differences in structural properties motivate the need to develop new channel structure-aware beam alignment techniques for short range channels. In this paper, we modeled the structure in short range mmWave channels by splitting the channel into several subchannels. We showed how exploiting subchannel correlation in addition to the sparse representation of each subchannel, can result in better beam alignment when compared to standard CS algorithms. We have also developed a low complexity message passing algorithm that ensures faithfulness of the solution to the problem geometry. Our results indicate that the algorithms based on dynamic CS and message passing are promising candidates to exploit structure in short range mmWave systems. Investigating the impact of antenna polarization and incorporating frequency selectiveness of channels are interesting research directions. In our future work, we will design efficient CS matrices for geometry aided-message passing. 
\vspace{-2mm}
\bibliographystyle{IEEEtran}
\bibliography{refs}

\begin{thebibliography}{10}
\providecommand{\url}[1]{#1}
\csname url@samestyle\endcsname
\providecommand{\newblock}{\relax}
\providecommand{\bibinfo}[2]{#2}
\providecommand{\BIBentrySTDinterwordspacing}{\spaceskip=0pt\relax}
\providecommand{\BIBentryALTinterwordstretchfactor}{4}
\providecommand{\BIBentryALTinterwordspacing}{\spaceskip=\fontdimen2\font plus
\BIBentryALTinterwordstretchfactor\fontdimen3\font minus
  \fontdimen4\font\relax}
\providecommand{\BIBforeignlanguage}[2]{{%
\expandafter\ifx\csname l@#1\endcsname\relax
\typeout{** WARNING: IEEEtran.bst: No hyphenation pattern has been}%
\typeout{** loaded for the language `#1'. Using the pattern for}%
\typeout{** the default language instead.}%
\else
\language=\csname l@#1\endcsname
\fi
#2}}
\providecommand{\BIBdecl}{\relax}
\BIBdecl

\bibitem{shortrange_spawc}
J.~Kaleva, N.~J. Myers, A.~T{\"o}lli, and R.~W. Heath~Jr, ``A geometry-aided
  message passing method for \text{AoA}-based short range \text{MIMO} channel
  estimation,'' in \emph{Proc. of the IEEE Intl. Conf. on Signal Proc. Adv. in
  Wireless Commun.}, 2019.

\bibitem{mmWavemotiv}
R.~C. Daniels and R.~W. Heath~Jr, ``60 \text{GHz} wireless communications:
  Emerging requirements and design recommendations,'' \emph{IEEE Vehicular
  technology magazine}, vol.~2, no.~3, 2007.

\bibitem{formfactor}
S.~Okasaka, R.~J. Weiler, W.~Keusgen, A.~Pudeyev, A.~Maltsev, I.~Karls, and
  K.~Sakaguchi, ``Proof-of-concept of a millimeter-wave integrated
  heterogeneous network for 5g cellular,'' \emph{Sensors}, vol.~16, no.~9, p.
  1362, 2016.

\bibitem{heathoverview}
R.~W. Heath, N.~Gonzalez-Prelcic, S.~Rangan, W.~Roh, and A.~M. Sayeed, ``An
  overview of signal processing techniques for millimeter wave \text{MIMO}
  systems,'' \emph{IEEE J. Sel. Topics Signal Process.}, vol.~10, no.~3, pp.
  436--453, 2016.

\bibitem{11ad}
T.~Nitsche, C.~Cordeiro, A.~B. Flores, E.~W. Knightly, E.~Perahia, and J.~C.
  Widmer, ``\text{IEEE} 802.11 ad: directional 60 \text{GHz} communication for
  multi-gigabit-per-second wi-fi,'' \emph{IEEE Commun. Mag.}, vol.~52, no.~12,
  pp. 132--141, 2014.

\bibitem{kiranchannel}
J.~Rodr{\'\i}guez-Fern{\'a}ndez, N.~Gonz{\'a}lez-Prelcic, K.~Venugopal, and
  R.~W. Heath~Jr, ``Frequency-domain compressive channel estimation for
  frequency-selective hybrid \text{mmWave} \text{MIMO} systems,'' \emph{IEEE
  Trans. Wireless Commun.}, vol.~17, no.~5, pp. 2946--2960, 2018.

\bibitem{lowrank1}
P.~A. Eliasi, S.~Rangan, and T.~S. Rappaport, ``Low-rank spatial channel
  estimation for millimeter wave cellular systems,'' \emph{IEEE Trans. on
  Wireless Commun.}, vol.~16, no.~5, pp. 2748--2759, 2017.

\bibitem{tsewicomm}
D.~Tse and P.~Viswanath, \emph{Fundamentals of wireless communication}.\hskip
  1em plus 0.5em minus 0.4em\relax Cambridge university press, 2005.

\bibitem{vaswani2010modified}
N.~Vaswani and W.~Lu, ``Modified-cs: Modifying compressive sensing for problems
  with partially known support,'' \emph{IEEE Trans. on Signal Process.},
  vol.~58, no.~9, pp. 4595--4607, 2010.

\bibitem{DCSAMP}
J.~Ziniel and P.~Schniter, ``Dynamic compressive sensing of time-varying
  signals via approximate message passing,'' \emph{IEEE Trans. on Signal
  Process.}, vol.~61, no.~21, pp. 5270--5284, 2013.

\bibitem{dynamiccs1}
L.~Lian, A.~Liu, and V.~K. Lau, ``Exploiting dynamic sparsity for downlink
  \text{FDD}-massive \text{MIMO} channel tracking,'' \emph{IEEE Trans. on
  Signal Process.}, vol.~68, no.~8, pp. 2007 -- 2021, 2019.

\bibitem{dynamiccs2}
A.~Liu, V.~K. Lau, and W.~Dai, ``Exploiting burst-sparsity in massive
  \text{MIMO} with partial channel support information,'' \emph{IEEE Trans. on
  Wireless Commun.}, vol.~15, no.~11, pp. 7820--7830, 2016.

\bibitem{dynamiccs3}
L.~Chen, A.~Liu, and X.~Yuan, ``Structured turbo compressed sensing for massive
  \text{MIMO} channel estimation using a markov prior,'' \emph{IEEE Trans. on
  Vehicular Tech.}, vol.~67, no.~5, pp. 4635--4639, 2018.

\bibitem{KiranBS}
K.~Venugopal, N.~Gonz{\'a}lez-Prelcic, and R.~W. Heath~Jr, ``Optimal
  frequency-flat precoding for frequency-selective millimeter wave channels,''
  \emph{submitted to IEEE Trans. on Wireless Commun.}, 2018.

\bibitem{FALP}
N.~J. Myers, A.~Mezghani, and R.~W. Heath, ``\text{FALP}: Fast beam alignment
  in \text{mmWave} systems with low-resolution phase shifters,'' \emph{arXiv
  preprint arXiv:1902.05714}, 2019.

\bibitem{rappwireless}
T.~S. Rappaport, ``Wireless communications: Principles and practice,'' 2002.

\bibitem{beamsparse}
J.~Wang, ``Beam codebook based beamforming protocol for multi-\text{Gbps}
  millimeter-wave \text{WPAN} systems,'' \emph{IEEE J. Sel. Areas in Commun.},
  vol.~27, no.~8, 2009.

\bibitem{swiftlink}
N.~J. Myers, A.~Mezghani, and R.~W. Heath, ``Swift-link: A compressive beam
  alignment algorithm for practical \text{mmWave} radios,'' \emph{IEEE Trans.
  on Signal Process.}, vol.~67, no.~4, pp. 1104--1119, 2019.

\bibitem{zadoffchu}
D.~Chu, ``Polyphase codes with good periodic correlation properties,''
  \emph{IEEE Trans. Inform. Theory}, vol.~18, no.~4, pp. 531--532, 1972.

\bibitem{ZCglobecom}
N.~J. Myers, A.~Mezghani, and R.~W. Heath, ``Spatial \text{Zadoff-Chu}
  modulation for rapid beam alignment in \text{mmWave} phased arrays,'' in
  \emph{Proc. of the IEEE Global Telecommun. Conf. (GLOBECOM)}, 2018.

\bibitem{ZCCS_recent}
C.-R. Tsai and A.-Y. Wu, ``Structured random compressed channel sensing for
  millimeter-wave large-scale antenna systems,'' \emph{IEEE Trans. on Signal
  Process.}, vol.~66, no.~19, pp. 5096--5110, 2018.

\bibitem{AMP}
D.~L. Donoho, A.~Maleki, and A.~Montanari, ``Message-passing algorithms for
  compressed sensing,'' \emph{Proceedings of the National Academy of Sciences},
  vol. 106, no.~45, pp. 18\,914--18\,919, 2009.

\bibitem{Proxilink}
N.~J. Myers, ``Beam alignment in short range mmwave systems,''
  \url{https://github.com/nitinjmyers}, 2019.

\bibitem{MMSE}
R.~W. Heath~Jr and A.~Lozano, ``Foundations of \text{MIMO}
  communication.''\hskip 1em plus 0.5em minus 0.4em\relax Cambridge University
  Press, 2018.

\bibitem{cycles_avoid}
M.~Karimi and A.~H. Banihashemi, ``Message-passing algorithms for counting
  short cycles in a graph,'' \emph{IEEE Trans. on Commun.}, vol.~61, no.~2, pp.
  485--495, 2013.

\bibitem{roomchannelmodel}
E.~Torkildson, U.~Madhow, and M.~Rodwell, ``Indoor millimeter wave \text{MIMO}:
  Feasibility and performance,'' \emph{IEEE Trans. on Wireless Commun.},
  vol.~10, no.~12, pp. 4150--4160, 2011.

\bibitem{slezak2018empirical}
C.~Slezak, V.~Semkin, S.~Andreev, Y.~Koucheryavy, and S.~Rangan, ``Empirical
  effects of dynamic human-body blockage in 60 \text{GHz} communications,''
  \emph{IEEE Commun. Mag.}, vol.~56, no.~12, pp. 60--66, 2018.

\bibitem{carpetrefl}
K.~Sato, H.~Kozima, H.~Masuzawa, T.~Manabe, T.~Ihara, Y.~Kasashima, and
  K.~Yamaki, ``Measurements of reflection characteristics and refractive
  indices of interior construction materials in millimeter-wave bands,'' in
  \emph{IEEE Vehicular Tech. Conf.}, vol.~1, 1995, pp. 449--453.

\bibitem{EMGAMP}
J.~P. Vila and P.~Schniter, ``Expectation-maximization \text{Gaussian}-mixture
  approximate message passing,'' \emph{IEEE Trans. Signal Process.}, vol.~61,
  no.~19, pp. 4658--4672, 2013.

\end{thebibliography}
\end{document}